\documentclass[a4paper,12pt,leqno, twoside]{article}

\usepackage{paper}

\hypersetup{pdftitle={Paper Example}}
\graphicspath{{./graphs/}{./graphs/lalonde/}{./graphs/irs}{./graphs/tables}}%

\usepackage{afterpage}

\usepackage[flushmargin]{footmisc}
\usepackage{tgpagella} 

\usepackage{lettrine}

\usepackage{indentfirst}
\usepackage{caption}
\usepackage{fancyhdr}
\newcommand{\indep}{\perp\!\!\!\perp}

\usepackage[colorinlistoftodos]{todonotes}
\usepackage{placeins}

\newcommand{\bib}{main.bib}

\begin{document}


\newpage
\setcounter{page}{1}
\thispagestyle{empty} 

\renewcommand{\thefootnote}{\fnsymbol{footnote}}

\begin{flushleft}
\doublespacing
\vspace*{3em}
{\LARGE \textbf{Comparing Experimental and Nonexperimental Methods: What Lessons Have We Learned Four Decades After LaLonde (1986)?}}\\
\vspace*{3.5em}


{\Large Guido W. Imbens and Yiqing Xu%
\footnote{\emph{Guido W. Imbens is an Applied Econometrics Professor and a Professor of Economics at the Graduate School of Business and the Department of Economics, Stanford University, Stanford, California. He is also a Research Associate, National Bureau of Economic Research, Cambridge, Massachusetts. Yiqing Xu is an Assistant Professor in Political Science at the Department of Political Science and a W. Glenn Campbell and Rita Ricardo-Campbell National Fellow at the Hoover Institution at Stanford University, Stanford, California. Their email addresses are imbens@stanford.edu and yiqingxu@stanford.edu.}\bigskip\\ 
For supplementary materials such as appendices, datasets, tutorials, and author disclosure statements, see the article page at https://doi.org/XXXX}
}
\vspace*{4.5em}
\end{flushleft}

\renewcommand{\thefootnote}{\arabic{footnote}}
\setcounter{footnote}{1}

\lettrine[lines=3]{I}{n} 1986, Robert LaLonde published a paper based on part of his PhD thesis  \citep{LaLonde}, which has had profound impact on both the methodological and empirical literatures on estimating causal effects. As of May 2025, this paper has been cited over 3,000 times, a number that only partially reflects its tremendous influence on the field of causal inference and the credibility revolution \citep{angrist2010}. 

The context for LaLonde's (1986) paper was the National Supported Work demonstration program. This program targeted individuals with extremely poor employment prospects: for females, recipients of Aid to Families with Dependent Children; for males, ex-drug addicts, ex-criminal offenders, and high school dropouts. Although about 10,000 people participated across 15 cities, the evaluation focused on approximately 6,600 participants in 10 cities. Approximately half were randomly assigned to a control group, while the other half received up to 12 months of training in small groups, with a supervisor and a counselor, followed by efforts to place them with outside employers. According to the Manpower Development Research Corporation (MDRC), which operated the program and conducted the evaluation, the National Supported Work program substantially increased 1978 earnings for female participants, particularly those without prior job experience; in contrast, the effects for male participants were smaller and, for some subgroups, essentially nonexistent (MDRC, \citeyear{mdrc1980}).

LaLonde used this experimental evaluation to address a broader methodological question: whether the state-of-the-art nonexperimental evaluation methods of that time could replicate experimental benchmarks---that is, estimates obtained from randomized controlled trials. By nonexperimental methods, we mean statistical approaches that estimate a program’s impact using naturally occurring variation in treatment assignment, without randomization or experimental control. LaLonde's conclusion was sharply negative regarding the credibility of the nonexperimental methods he examined. He wrote \cite[Abstract, p. 604]{LaLonde}:
\begin{quote}
This comparison shows that many of the econometric procedures do not replicate the experimentally determined results, and it suggests that researchers should be aware of the potential for specification errors in other nonexperimental evaluations.
\end{quote}
and concluded
 \cite[Conclusion, p. 617]{LaLonde}:
\begin{quote}
[P]olicymakers should be aware that the available nonexperimental evaluations of employment and training programs may contain large and unknown biases resulting from specification errors. 
\end{quote}

\noindent Around the time LaLonde’s paper was published, several other studies made similar points about the credibility of nonexperimental methods, including work by LaLonde’s thesis advisers \citet{ashenfelter1985using}, as well as \citet{fraker1987adequacy} and \citet{heckman1987we}. 

One reason LaLonde’s study became so influential is that the full data underlying the original analysis later became publicly available. As part of a research project initiated in a 1996 graduate class taught by Imbens and Rubin at Harvard, \citet{dehejiawahba, dehejia2002propensity} retrieved LaLonde’s original male subsample data—stored on tapes—by locating a tape reader capable of reading the files. The data are now publicly available on Dehejia’s website (\url{https://users.nber.org/~rdehejia/data/.nswdata2.html}). We refer to a subsample of these data that is most commonly used in the literature as the LaLonde-Dehejia-Wahba (LDW) data. More recently, \citet{calonico2017women} also reconstructed the female samples. These public datasets are highly valuable for both teaching and future research and are provided alongside this paper, with code to replicate the estimates reported here.

The methodological literature on nonexperimental evaluation has advanced significantly. Nearly four decades after LaLonde’s paper, the answer to his original question—whether nonexperimental methods can successfully replicate experimental benchmarks—is more nuanced than his initial conclusion: sometimes they can, and we now have better tools to help assess when they are likely to succeed.

Here are five key lessons that emerge from this literature since the publication of \citet{LaLonde}. First, the unconfoundedness assumption has become central to modern methods for nonexperimental data. Unconfoundedness means that, conditional on pre-treatment variables or covariates, treatment is as if randomly assigned. This assumption elegantly separates the underlying identification assumptions from functional form considerations.  It emphasizes \emph{design}—how treatment is assigned—over full specification of the data-generating process.

Second, inspecting and ensuring overlap is critical in the process of estimating treatment effects. Overlap means that the covariate distributions in treatment and control groups have common support. Improved overlap reduces sensitivity to the choice of estimation strategy and improves the robustness of estimates.

Third, the use of propensity scores, defined as the probability of receiving treatment given observed covariates, has become widespread. The propensity score, introduced in \citet{rosenbaum1983central},  had only just entered the economics literature when \citet{LaLonde} was writing his thesis. Since then, propensity scores have become a core component of many analyses  that rely on unconfoundedness, including for assessing overlap, but also for estimation.

Fourth, researchers have increasingly taken heterogeneity in treatment effects seriously. This includes both methods for estimating average treatment effects in the presence of heterogeneity, but also going beyond average treatment effects to examine estimates of treatment effect heterogeneity, such as conditional treatment effects and quantile treatment effects, {\it e.g.}, \citet{wager2018estimation}. These approaches provide a deeper understanding of how the treatment impacts different individuals or segments of the outcome distribution.

Finally, validation exercises—particularly placebo tests—are essential for assessing key assumptions and evaluating the credibility of causal claims. In a placebo test, researchers use an outcome that should not be affected by the treatment. For example, \citet{LaLonde} used 1975 earnings as a placebo outcome, since it predates the training program. A non-zero estimated effect on the placebo outcome indicates potential unobserved confounding. Placebo tests are important diagnostic tools in modern nonexperimental research since the credibility revolution \citep{angrist2010}.%
\footnote{We focus on five issues specific to the setting studied in \citet{LaLonde}: the evaluation of an individual-level intervention using detailed background information on participants. While the broader literature on causal inference has expanded substantially over the past four decades \citep[see][for an illustration of the changes in methodologies used over this period]{currie2020technology}, we do not attempt to cover that broader landscape here. For more comprehensive overviews, see surveys such as \citet{imbens2009recent} and \citet{abadie2018econometric}, as well as a number of textbooks \citep{angrist2008mostly, imbens2015causal, cunningham2018causal, huntington2021effect, huber2023causal, ding2024first, wager2024, chernozhukov2024applied}.}

To illustrate these lessons, we reexamine the original and reconstructed LaLonde datasets. We show that, once overlap is ensured, various modern methods yield similar estimates. However, these estimates lack a causal interpretation if unconfoundedness is violated. Validation exercises such as placebo tests are critical for assessing the credibility of unconfoundedness. For the LaLonde datasets, placebo estimates generally do \emph{not} support the unconfoundedness assumption.

\section{LaLonde's Findings}\label{findings}

We begin by describing the data used in \citet{LaLonde} and outlining the main econometric approaches and results presented in the paper. We then introduce the LaLonde-Dehejia-Wahba data and examine some of the immediate methodological responses to LaLonde’s findings.

\subsection{LaLonde's Data}

Although the MDRC evaluation included 6,600 individuals, the experimental samples LaLonde used to evaluate nonexperimental methods were much smaller. For males, the sample included 722 individuals (297 treated and 425 controls), and for females, 1,158 individuals (600 treated and 585 controls). This substantial reduction in sample size stemmed from three main factors. First, many participants failed to complete follow-up interviews needed to collect post-program earnings data. Second, due to budget constraints, the MDRC team randomly selected subsamples for the 27- and 36-month follow-up interviews. Third, LaLonde excluded male participants who entered the program before January 1976 or were still enrolled in January 1978.

To conduct a nonexperimental analysis of the effects of the training program, LaLonde combined the experimental treatment group with comparison groups drawn from external, nonexperimental datasets used as controls. For both female and male participants, he used two main sources. The first, CPS-SSA-1, is based on Westat's Matched Current Population Survey–Social Security Administration File and includes all females or males under age 55 who met Westat's eligibility criteria. The second, PSID-1, comes from the Panel Study of Income Dynamics and includes all female or male household heads under age 55 from 1975 to 1978 for males and 1979 for females, excluding those who identified as retired in 1975. The age cutoff was intended to improve comparability between the comparison group and the experimental sample.


\begin{table}[!ht]
\begin{minipage}[c]{1\textwidth}
\caption{\\Descriptive Statistics: LaLonde and Lalonde-Dehejia-Wahba  Male Samples}\label{tb:samples}
\vspace{-0.5em}\includegraphics[width=\linewidth]{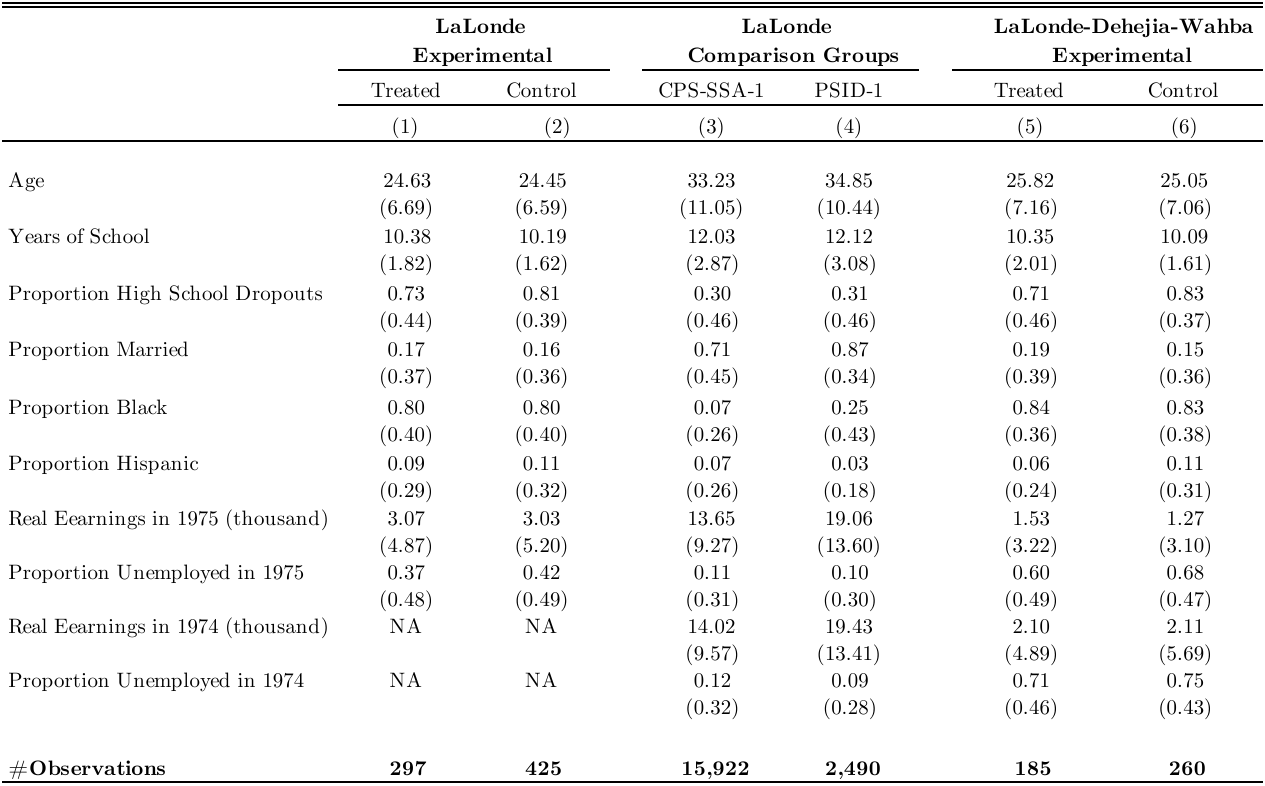}
{\footnotesize\textbf{\textit{Note:}} Standard deviations are shown in parentheses. The experimental data were drawn from the National Supported Work program. Tables 3, 5, and 6 in \citet{LaLonde} use the data summarized in columns 1–4, while \citet{dehejiawahba} primarily rely on the data in columns 3–6.}
\end{minipage}%
\end{table}

Table~\ref{tb:samples} columns 1–4 present summary statistics for the male samples in the experimental data and the comparison groups used in \citet{LaLonde}. Respondents in both the CPS and PSID comparison groups were, on average, older, more educated, substantially less likely to be high school dropouts, more likely to be married, and less likely to be Black or Hispanic than participants in the National Supported Work experiment. They also had much higher earnings in the years before the program. For example, average 1975 earnings were \$3,000 in the experimental data, compared to \$14,000 in the CPS-SSA-1 sample and \$19,000 in the PSID-1 sample. Notably, the comparison groups are much larger: the male CPS-SSA-1 sample includes 15,922 observations, and the male PSID-1 sample contains 2,490 observations. To improve comparability with the experimental sample, LaLonde dropped from these datasets individuals based on employment status, time of survey, and poverty status, creating four additional, smaller comparison groups. 

Columns 5 and 6 of Table~\ref{tb:samples} present summary statistics for the experimental treated and control units in the LaLonde-Dehejia-Wahba data (discussed further below), a subset of the original LaLonde male sample selected for containing information on  1974 earnings. Our reanalysis focuses on this dataset because it is widely used in the post-LaLonde methodological literature and includes two pretreatment outcomes—earnings in 1974 and 1975—which enable adjustment for longer earnings histories and placebo analyses. Results for the original LaLonde male samples and reconstructed female samples are reported in the online appendix.

\subsection{Econometric Approaches in LaLonde (1986)}

To estimate the causal effect of the National Supported Work program on 1978 earnings using both experimental and nonexperimental data, LaLonde employed a variety of models that fall into two broad categories: regression methods, in which earnings serve as the outcome (referred to as the ``earnings equation''), and selection models that also include a ``participation equation,'' where the outcome is program participation. In LaLonde's paper, Tables 4 and 5 present the regression results, while Table 6 reports the selection model results. These tables are reproduced in the appendix.

LaLonde reported training effect estimates for both female and male participants using seven estimators across six comparison groups: two large groups described in Table~\ref{tb:samples} and the aforementioned four subgroups defined through additional selection criteria to improve overlap with the National Supported Work demonstration sample. All regression models were linear and implicitly assumed constant treatment effects. Of the seven models, two are simple regressions estimated without or with controls for age, education, and race (but notably excluding 1975 earnings). Two models use a difference-in-differences estimator, replacing the outcome with the change in earnings between 1975 and 1978, estimated with and without controlling for age. Two more use a quasi-difference-in-differences specification, including 1975 earnings on the right-hand side to account for transitory shocks—known as the ``Ashenfelter dip'' \citep{ashenfelter1978estimating}—again estimated with and without controlling for age. The final specification includes all available pretreatment covariates, including 1975 earnings, 1975 unemployment status, and marital status.

In an additional analysis in the spirit of the modern causal literature, LaLonde also conducted placebo tests using 1975 earnings as the outcome. Because the training occurred after 1975, the true treatment effect on this placebo outcome is zero. And indeed, the placebo estimates from the linear regression approach on the experimental sample were close to zero.

LaLonde emphasized several key findings from the linear regression results. First, using the experimental data, all seven estimators produced similar estimates: around \$851 for female participants and \$886 for male participants. Second, when using nonexperimental comparison groups, the estimates diverged sharply from these benchmarks, often yielding large, negative values in both female and male samples, with modest standard errors. Third, the estimates from nonexperimental data varied widely across specifications, and goodness-of-fit tests offered little guidance for selecting models that aligned with the experimental results. Taken together, these findings led LaLonde to conclude that the regression adjustment methods commonly used at the time were not credible when applied to nonexperimental data.

In addition to the regression models, which rely on exogeneity of the treatment indicator, LaLonde also presented results from selection models that allow for potential endogeneity. These models use the two-step estimator proposed by \citet{heckman1978}, which permits correlation between the error terms in the earnings and participation equations. Identification in this framework relies either on exclusion restrictions—variables that appear in the participation equation but not in the earnings equation—or on functional form and distributional assumptions  (linearity and joint normality of the error terms). For both female and male samples, LaLonde used three comparison groups (the experimental controls, CPS-SSA-1, and PSID-1) and estimated four specifications for females and three for males. Each specification used a different set of excluded variables. {\it A priori}, no single specification was clearly more defensible based on economic or econometric reasoning.

LaLonde found that the selection model estimates using the experimental data remained close to \$851 for females and \$886 for males. However, those using nonexperimental data again varied substantially and deviated from the experimental benchmarks. He concluded that, although the two-step procedure brought the estimates somewhat closer to the benchmarks, it still yielded a ``considerable range of imprecise estimates'' \cite[p. 617]{LaLonde}.

\subsection{The LaLonde-Dehejia-Wahba Data}

The data reconstructed by \citet{dehejiawahba} focused exclusively on male participants, noting that ``estimates for this group were the most sensitive to functional-form specification'' \cite[p. 1054]{dehejiawahba}. They constructed a subsample from LaLonde’s original data consisting of individuals with available information on 1974 earnings and unemployment status. They argued that this subsample remained a valid experimental sample because it was constructed solely using pretreatment information—such as month of assignment and employment history—thereby preserving the orthogonality of treatment assignment with respect to observed and unobserved characteristics. Notably, this subsample includes only 62 percent of the original treatment group in LaLonde’s study. For nonexperimental controls, they used subsets of the same datasets LaLonde employed, restricted to units with 1974 earnings and unemployment data. This collection, now widely known as the LaLonde-Dehejia-Wahba (LDW) data, has become a standard benchmark in the causal inference literature. In particular, most methodological studies focus on the version combining the experimental treated units with CPS-SSA-1 controls. For brevity, we refer to the LaLonde-Dehejia-Wahba experimental sample as LDW-Experimental, and the samples combining these treated units with CPS-SSA-1 and PSID-1 controls as LDW-CPS and LDW-PSID, respectively.

As noted earlier, columns 5 and 6 of Table~\ref{tb:samples} report summary statistics for the LDW-Experimental sample. Compared to LaLonde's original male sample, participants in this subsample had higher unemployment rates and lower average earnings in 1975. The inclusion of 1974 earnings---available only in the LaLonde-Dehejia-Wahba data---further suggests that many participants faced long-term unemployment. These differences may help explain why the estimated training effect in this sample (\$1,794) is more than twice that of LaLonde's original male sample (\$886).

\subsection{Subsequent Literature}

The publication of \citet{LaLonde} sparked a debate in the applied econometrics literature. For example, \citet{heckman1989choosing} responded to LaLonde’s critique of nonexperimental evaluation methods by advocating the use of specification tests to rule out particularly poor estimators. However, this approach did not offer a clear way to distinguish among the many estimators that fit the data reasonably well but rely on different identification assumptions. As a result, it gained limited traction in subsequent work.

In contrast, \citet{dehejiawahba} proposed an alternative, more flexible estimators  to address LaLonde’s challenge. Their proposals included approaches based on propensity score stratification and matching. Their estimates closely matched the experimental benchmark, leading them to conclude, in sharp contrast to LaLonde:
\begin{quote}
``[T]he estimates of the training effect for LaLonde's ... dataset are close to the benchmark experimental estimates and are robust to the specification of the comparison group and to the functional form used to estimate the propensity score. ... our methods succeed for a transparent reason: They use only the subset of the comparison group that is comparable to the treatment group, and discard the complement.'' \cite[p. 1062]{dehejiawahba}.
\end{quote}
The contrast between the conclusions of \citet{dehejiawahba} and \citet{LaLonde} led to a wave of methodological research aimed at probing and reconciling these findings. We turn to these developments in the next section.

\section{Methodological Improvements since \citet{LaLonde}}

To facilitate discussion of methodological advances since \citet{LaLonde}, we begin with the ``potential outcomes framework,'' then introduce the main causal estimand and a closely related statistical estimand. We then outline the key assumptions required for nonexperimental data to identify the causal estimand.

\subsection{Causal and Statistical Estimands}

Over the past four decades, the applied econometrics literature has made significant progress—most notably in clarifying estimands (what is being estimated) and the assumptions required for identification (how to use data to recover them). Once these assumptions are clearly stated, researchers can assess how plausible they are.

In our discussion, we adopt the potential outcome framework, originally developed by Jerzy Neyman in the context of randomized experiments \citep{neyman1923} and extended to nonexperimental settings by Donald Rubin \citep{rubin1974estimating, rubin2006matched, imbens2015causal}. For each individual $i$, two potential outcomes are postulated: $Y_i(0)$ and $Y_i(1)$. In the LaLonde setting, they represent the individual's 1978 earnings had they not participated in the program and had they participated, respectively. The causal effect for individual $i$ is the difference $Y_i(1) - Y_i(0)$. Let $W_i$ denote the binary treatment indicator, equal to 1 if the individual participated in the program and 0 otherwise. The observed outcome is linked to the potential outcomes by the identity $Y_i = (1 - W_i) Y_i(0) + W_i Y_i(1)$. We also observe pretreatment characteristics or covariates $X_i$. In the LaLonde study, these include age, years of schooling, high school dropout status, marital status, and indicators for Black and Hispanic backgrounds.\footnote{The modern methodological literature has paid particular attention to settings where the vector of pretreatment variables is high-dimensional.} Following \citet{dehejiawahba}, we may augment this vector to include 1974 and 1975 earnings and indicators for zero earnings (unemployment) in each year.

Our primary goal is to estimate the average treatment effect on the treated (ATT), 
\[
ATT = \mathbb{E}[Y_i(1) - Y_i(0) \mid W_i = 1]\  =  \ \mathbb{E}[Y_i \mid W_i = 1]  - \mathbb{E}[Y_i(0) \mid W_i = 1].
\]
ATT represents the individual-level causal effect averaged over for those who actually received the treatment.\footnote{For simplicity, here we do not distinguish between a finite study population and a super-population.} In LaLonde's context, it means the average treatment effect of the job training program on individuals who actually participated. Since $Y_i(1) = Y_i$ for treated individuals (individuals with $W_i = 1$), the second equality follows. The first term on the right-hand side, $\mathbb{E}[Y_i \mid W_i = 1]$, can be directly estimated from the observed data. The second term, $\mathbb{E}[Y_i(0) \mid W_i = 1]$, cannot, because it involves \emph{counterfactual outcomes}—in this case, what participants' earnings would have been, had they not participated in the program.  

Most analyses of the LaLonde data that explicitly allow for heterogeneous treatment effects focus on the ATT, as it makes little sense to estimate or even contemplate the effect of the program for nonparticipants with stable jobs and high earnings. In other contexts, researchers may be interested in the average treatment effect (ATE), which captures the average effect across the entire population while the ATT focuses specifically on the treated group. \citet{LaLonde} did not explicitly distinguish between different estimands, as his analysis did not consider treatment effect heterogeneity.

Because we cannot observe the counterfactual outcomes, we cannot directly estimate the ATT. This is the core of what \citet{holland1986statistics} termed the ``fundamental problem of causal inference.'' To make progress, we can instead estimate the covariate-adjusted difference in average outcomes between treated and control groups:
\[ 
\text{(Covariate-adjusted difference)}\quad \mathbb{E}[Y_i \mid W_i = 1] - \mathbb{E}\left\{\mathbb{E}[Y_i|W_i=0,X_i]\mid W_i=1\right\}.
\]
We refer to this as a \emph{statistical estimand}, as opposed to a \emph{causal estimand}, because it can be estimated from observed data when overlap holds, an assumption we return to below. In LaLonde's context, the statistical estimand represents the difference between the average 1978 earnings of individuals who participated in the program and the average 1978 earnings of nonexperimental individuals with similar observed characteristics.

Unlike a statistical estimand, the ATT is a \emph{causal estimand} because it compares potential outcomes for the same individuals—specifically, those who participated in the program. These two estimands are not necessarily equal. The counterfactual mean for the treated individuals, $\mathbb{E}[Y_i(0) \mid W_i = 1]$, may differ from $\mathbb{E}\left\{\mathbb{E}[Y_i \mid W_i = 0, X_i] \mid W_i = 1\right\}$, a weighted average of observed outcomes for individuals in the nonexperimental control group with similar characteristics. For the statistical estimand to recover the ATT, a key assumption—unconfoundedness—must hold. That is, if unconfoundedness holds, we can approximate the average untreated outcome for the treated individuals using outcomes from control individuals of similar observed characteristics.

A substantial body of subsequent research has focused on improving methods for estimating the covariate-adjusted difference, particularly in high-dimensional settings. Many recent approaches draw on machine learning techniques to flexibly estimate the relationships between covariates, treatment, and outcomes. Formal results often rely on additional regularity conditions, such as the smoothness of conditional means and propensity scores. For formal treatments, see the references in \citet{imbens2009recent} and \citet{abadie2018econometric}. The methodological advances in this literature apply well beyond the LaLonde setting.

Our focus here is on the plausibility of the key assumptions. In particular, if unconfoundedness does not hold, we may still be able to robustly and precisely estimate the covariate-adjusted difference, but it cannot be interpreted as a causal effect and may have little substantive relevance.

\subsection{Unconfoundedness}

The unconfoundedness assumption was first explicitly introduced by \citet{rubin1978bayesian} as part of the ``ignorable treatment assignment'' concept, see also \citet{rosenbaum1983central}. If unconfoundedness holds, then treatment assignment can be viewed as effectively random once differences in observed covariates between the treatment and control groups are adjusted for. This assumption plays a central role in identifying causal effects from nonexperimental data. In the context of estimating the ATT, unconfoundedness is formally stated as:
\[
\text{(Unconfoundedness)}\qquad W_i\ \indep \ Y_i(0)\ |\ X_i\ ,
\]
which means that treatment status is conditionally independent of  the control potential outcome given the covariates. In LaLonde’s context, unconfoundedness means that once we account for observable characteristics (including age, education, race, marital status, and prior earnings), whether someone was in the experimental treatment group or nonexperimental comparison group provides no additional information about what their 1978 earnings would have been had they not participated in the program.

Unconfoundedness is also referred to as \emph{conditional independence} \citep{lechner1999earnings, lechner2002program}, or informally as  {\it exogeneity} \citep{imbens2004}, or {selection on observables} \citep{barnow1980issues}. However, its definition departs from traditional econometric definitions of exogeneity which are expressed in terms of residuals and specific functional forms. By contrast, the formal statement of unconfoundedness  avoids functional form assumptions, focusing instead on the treatment assignment mechanism. It allows researchers to separate the essence of the identification assumptions from functional form considerations, emphasizing the role of design rather than the full specification of the data-generating process.

A key result from \citet{rosenbaum1983central} shows that unconfoundedness implies that conditioning on the scalar propensity score is sufficient to remove bias from covariate imbalance. This dimensionality reduction—from the full vector $X_i$ to a single index—has made the propensity score a cornerstone of many modern estimators.

When the parametric model for the conditional expectation of the outcome—such as the earnings equation used in \citet{LaLonde}—is correctly specified, unconfoundedness implies a zero conditional mean for the error term. Thus, the results reported in Tables 4 and 5 of \citet{LaLonde} can be viewed as relying on a combination of unconfoundedness and correct functional form assumptions. At the time, however, the nonparametric framing of the identifying assumptions and the emphasis on assignment mechanisms were not yet standard in applied work.

In practice, unconfoundedness is a strong assumption, and its plausibility depends heavily on the context. When the treatment assignment mechanism is poorly understood, this assumption may not be credible. Nonetheless,  often researchers can assess its plausibility through supplementary analyses. Tools such as placebo tests and sensitivity analyses can help probe the credibility and robustness of causal claims that rely on unconfoundedness. For a general discussion, see \citet{rosenbaum1983central} and \citet{imbens2004}.

In the next section, we illustrate the use of placebo tests to evaluate unconfoundedness. In the LaLonde setting, the ten covariates are clearly pretreatment variables and should be included in any adjustment strategy. In other settings, however, whether a given covariate should be adjusted for is less obvious. \citet{rosenbaum1984consequences} cautions against adjusting for post-treatment variables, and \citet{cinelli2022crash} offer guidance on selecting from among valid pretreatment covariates for causal inference.

\subsection{Overlap and Balance}\label{s:overlap}

To identify the ATT—and to ensure that the statistical estimand is properly defined—we require an overlap assumption, which states that the propensity score is strictly less than one:
\[
\text{(Overlap)}\qquad\Pr(W_i = 1 \mid X_i) < 1.
\]
In LaLonde's context, this means that for any individual $i$ assigned to the treatment group with a covariate profile $X_i = x_0$, there must also be individuals in the nonexperimental comparison group with the same profile; otherwise, $\Pr(W_i = 1 \mid X_i = x_0) = 1$, violating the overlap assumption. Overlap ensures that the weighted average $\mathbb{E}\left\{\mathbb{E}[Y_i \mid W_i = 0, X_i] \mid W_i = 1\right\}$, the second term in the statistical estimand, is well-defined. When overlap is violated, it can be restored by trimming the treatment group—that is, by removing treated units whose covariate profiles are not represented in the control group.\footnote{If the target parameter is the ATE, a stronger version of the overlap assumption is needed: the propensity score must lie strictly between zero and one for all units.}

Overlap implies that treated and control units must share common support in their covariate distributions. Without overlap—for instance, when some covariate values appear in the treatment group but not in the control group—it becomes difficult to make credible comparisons because estimates rely on extrapolation. Overlap is conceptually distinct from balance, which refers to the similarity in covariate distributions across groups. In a randomized control trials overlap holds by design, and balance is achieved in expectation. In that case, balance can be further improved through stratification or post-stratification. In nonexperimental settings, ensuring overlap is a key condition for credible estimation of causal effects.

Overlap is especially important when researchers do not wish to impose strong functional form assumptions on the conditional means of potential outcomes or on the structure of treatment effect heterogeneity. When the number of covariates is small, overlap can be assessed by examining marginal or joint covariate distributions across treatment groups. But this becomes impractical in high-dimensional settings. In such cases, it is more effective to inspect the distribution of estimated propensity scores across treated and control groups. Lack of overlap in covariate distributions implies, and is implied by, a lack of overlap in propensity score distributions.

LaLonde did not explicitly discuss overlap, nor did he assess it beyond reporting covariate means by treatment status. Both the regression and selection models he used rely on correct functional form assumptions, which permit interpolation or extrapolation of treatment effects even when treated and control units differ substantially in their covariates—thus bypassing the need for overlap. Still, LaLonde clearly recognized the potential problem: to improve comparability, he trimmed the comparison groups based on ``characteristics [that] are consistent with some of the eligibility criteria used to admit applicants into the NSW program'' \cite[p. 611]{LaLonde}. However, by modern standards, his trimming procedures—such as removing all men working in March 1976 in one subset (CPS-SSA-2) or further excluding unemployed respondents with 1975 incomes above the poverty line in another (CPS-SSA-3)—are {\it ad hoc} and certainly do not guarantee overlap on all relevant covariates.

Over the past four decades, researchers have developed more systematic approaches for improving overlap, often using the propensity score. These methods vary depending on whether the goal is simply to ensure overlap or to further improve covariate balance.

Improving overlap or balance typically requires dropping some units from the sample. Although this reduces the sample size, by ensuring better balance it may improve precision for estimates of the average treatment effect. In addition, the gain in robustness and reduction in bias often outweigh any loss in precision. In practice, any increase in variance, even  from substantial trimming is typically modest.\footnote{For example, suppose we have a sample with $N_{tr}$ treated units and $N_{co}$ control units. Under homoskedasticity and random assignment, the variance of the difference-in-means estimator is $\sigma^2(1/N_{tr} + 1/N_{co})$. In the LDW-CPS sample, starting with $N_{tr} = 185$ and $N_{co} = 15,922$, dropping 15,737 control units—a 99\% reduction—raises the standard error by only about 30\% in the ``best-case scenario,'' which assumes no bias from including the additional controls.}

Several specific trimming strategies have been proposed. For instance, focusing on overlap for ATT estimation, \citet{dehejiawahba} dropped control units with estimated propensity scores below the minimum observed in the treated group. \citet{crump2009dealing} proposed a more aggressive approach, selecting subsamples that minimize the variance of ATE estimates and recommending trimming units with propensity scores outside the $[.1, .9]$ interval. \citet{crump2006moving} and \citet{li2018balancing} proposed improving balance through propensity score weighting, introducing ``overlap weights'' proportional to the product of the propensity score and one minus the propensity score. Another effective strategy—especially when targeting the ATT—is to match each treated unit to a control unit with a similar estimated propensity score. This approach not only ensures overlap but also tends to improve balance across the covariate distributions.

In practice, overlap—like unconfoundedness—is essential for credible estimation. In settings with poor overlap, such as the LaLonde nonexperimental samples, trimming the sample to ensure overlap is often more important than the choice of estimation method.

\subsection{Estimation Given Unconfoundedness and Overlap}\label{s:estimation}

All estimators in \citet{LaLonde} are linear in the covariates. Since then, a wide range of more flexible methods have been proposed to estimate average causal effects under unconfoundedness and overlap. These approaches can be broadly categorized into three groups: (i) outcome modeling, including linear regressions, (ii) methods that directly adjust for covariate imbalance, including those based on propensity scores, and (iii) doubly robust methods.

First, outcome modeling remains the most widely used approach among applied researchers. It typically involves regressing the outcome on the treatment indicator and covariates (usually, the level terms)—what LaLonde called the earnings equation. This method assumes linearity in covariates and constant treatment effects. A modest relaxation of this approach involves estimating two separate linear regressions for the treated and control groups, sometimes referred to as the Oaxaca-Blinder estimator \citep{kline2011oaxaca}. More flexible alternatives include semiparametric and nonparametric methods to model the conditional means of the potential outcomes \citep[{\it e.g.},][]{heckman1997matching, athey2019generalized}.

The second group of methods focuses on directly adjusting for covariate imbalance between the treatment and control groups. This includes blocking on covariates ({\it i.e.}, grouping units with similar characteristics and comparing outcomes within groups, and then aggregating), covariate matching \citep[e.g.,][]{abadie2006, abadie2008failure, abadie2011bias, abadie2016matching, diamond2013genetic, rubin2006matched, imbens2015}, and weighting methods to achieve covariate balance \citep[{for example},][]{hirano2003efficient, hainmueller, zubizarreta2023handbook, zubizarreta2015stable}.

Covariate matching is a nonparametric method that avoids imposing modeling assumptions. However, it suffers from the curse of dimensionality when many covariates are present, making it prone to large biases \citep{abadie2006}. In such settings, adjusting for the estimated propensity score is often more practical. This can be implemented through blocking and matching \citep[{\it e.g.},][]{dehejiawahba, abadie2011bias} or through inverse propensity weighting (IPW). IPW reweights observations by the inverse of their estimated propensity score, creating a pseudo-population in which treatment assignment is uncorrelated to observed covariates. \citet{hirano2003efficient} show that the Hájek variant of the IPW estimator can achieve the semiparametric efficiency bound even when the propensity score is estimated nonparametrically. In the LaLonde setting, the IPW estimator reweights the nonexperimental control group based on estimated propensity scores so that its covariate distribution closely approximates that of the experimental treated group.

Since covariate imbalance is the sole source of bias under unconfoundedness, scholars have developed methods to improve balance either by refining propensity score estimation or by bypassing it entirely. For instance, \citet{imai2014covariate} propose estimating a covariate-balancing propensity score using the generalized method of moments, whereas \citet{hainmueller} introduces entropy balancing to directly achieve balance in specified covariate moments. Entropy balancing can be viewed as an IPW estimator that implicitly relies on a correctly specified propensity score model, where the link function is logistic and the log-odds are linear in covariates \citep{zhao2017entropy}.

However, neither outcome modeling nor matching or weighting methods on their own are currently most favored in the methodological literature. Instead, hybrid approaches combine outcome modeling, such as regression, with techniques that address covariate imbalance, such as propensity score weighting, to leverage the strengths of both. Examples include regression within propensity score blocks \citep{rosenbaumrubin1983assessing, imbens2015}, matching followed by regression adjustment \citep{rubin1973use, abadie2011bias}, and methods that integrate weighting with regression \citep[{\it e.g.},][]{robins1994estimation, robins1995semiparametric}. These approaches are motivated by the fact that even if balancing or propensity score methods are consistent and efficient in large samples, combining them with outcome modeling can reduce small-sample bias and improve precision. For instance, in high-dimensional settings, the bias from a matching estimator due to remaining covariate imbalance may dominate the variance, and regression adjustment that accounts for this imbalance  can help reduce the bias \citep{abadie2011bias}.

\citet{robins2001comment} introduced the term \emph{double robustness}, a key concept for hybrid methods. They show that if either the propensity score or the outcome model is correctly specified, the augmented inverse propensity weighting (AIPW) estimator—which combines propensity score weighting and regression—is consistent. AIPW can be viewed as an outcome model augmented by a correction term: an IPW estimator applied to the residuals from the outcome model, rather than the raw outcomes. The double robustness property arises because, if the outcome model is correctly specified, the correction term has mean zero even if the propensity score model is misspecified; if the outcome model is misspecified, a correctly specified IPW component can debias the regression model. Beyond robustness, AIPW is also desirable for its efficiency: when both models are correctly specified, it achieves the semiparametric efficiency bound \citep{bang2005doubly}.


More recently, machine learning methods have become increasingly popular in applied causal inference \citep{van2011targeted, wager2017estimation, chernozhukov2017double, athey2018approximate, athey2019generalized}; see  \citet{athey2017state, athey2019machine} for reviews. These methods are particularly useful for estimating ``nuisance parameters''---such as propensity scores or conditional outcome means---that are not of direct interest but are are essential for identifying causal effects. In the LaLonde setting, the goal is to estimate the ATT on post-training wages, which requires modeling the propensity score and/or the conditional means of potential outcomes. Many modern estimators satisfy the ``Neyman orthogonality'' condition \citep{chernozhukov2017double}, which reduces the effect of small estimation errors in the nuisance parameters on estimates of the target parameter. In binary treatment settings like LaLonde’s, these estimators resemble the AIPW estimator introduced earlier, but with the nuisance parameters estimated via flexible machine learning methods instead of parametric models. \citet{chernozhukov2017double, chernozhukov2018double} emphasize that this property ensures valid inference even when using machine learning algorithms that converge more slowly than is required for estimators based on only estimating the conditional outcome distributions or the propensity score.

\subsection{Alternative Estimands and Heterogeneous Treatment Effects}

Much of the methodological and applied research has focused on estimating average causal effects, such as the ATT. However, other quantities may also be of interest. For example, researchers often seek to understand treatment effect heterogeneity among treated units, which can shed light on mechanisms, improve evaluations of effectiveness, and guide personalized policy design. For example, the MDRC team reported that the National Supported Work program had a large and positive effect on female participants, a significant impact on ex-addict male participants, a small and highly variable impact on ex-criminal male participants, and almost no effect on youth participants (MDRC, \citeyear{mdrc1980}). Econometrically, exploring treatment effect heterogeneity given observed characteristics involves estimating the conditional average treatment effects on the treated (CATT). Machine learning methods have been proposed to estimate CATT either nonparametrically or using low-dimensional representations, such as causal forests, while still permitting valid inference or error bounds \citep[{\it e.g.},][]{athey2016recursive, wagerathey, athey2019generalized}. 

Another important, though less frequently used, estimand is the quantile treatment effects—defined as the difference between quantiles of the treated and untreated potential outcome distributions, either for the population or for the treated group. Under unconfoundedness and overlap, the full marginal distributions of potential outcomes are identified, enabling identification of quantile treatment effects. \citet{firpo2007efficient} proposes a semiparametrically efficient inverse propensity weighting estimator for these quantities. 

\subsection{Validation through Placebo Analyses}\label{placebo}

While researchers can assess overlap using observed data, the unconfoundedness assumption is not directly testable. To evaluate the credibility of treatment effect estimates, the literature has developed two main strategies: placebo analyses and sensitivity analyses. Here, we focus on the former and relegate discussion of the latter to the online appendix.

Placebo analyses offer an indirect way to probe the plausibility of unconfoundedness. These analyses typically involve estimating a model similar to the main specification, but replacing the outcome variable with a pseudo-outcome—usually a variable known to be unaffected by the treatment. A common approach is to test for a treatment effect on a pretreatment variable, such as a lagged outcome, which should not be influenced by the treatment but may still correlate with unobserved confounders. Another variant tests the effect of a pseudo-treatment—often a proxy for the treatment—on the actual outcome. This strategy is often implemented using multiple control groups. For further discussion, see \citet{rosenbaum1987role}, \citet{imbens2015causal}, and \citet{imbens2015}.

Although the term ``placebo test'' was not yet in use in the mid-1980s, LaLonde conducted such an analysis. He regressed 1975 earnings, a pretreatment variable, on the treatment indicator and covariates (reported in columns 2 and 3 of his Tables 4 and 5). Using nonexperimental data, he found that many of the estimated effects were large, negative, and statistically significant, suggesting a violation of unconfoundedness. One limitation of the LaLonde data is that it contains only a single pretreatment outcome. In contrast, the LaLonde-Dehejia-Wahba subset of the LaLonde data allow for placebo analyses that condition on an earlier pretreatment variable, 1974 earnings, when testing for associations between treatment and 1975 earnings. In general, when multiple pretreatment periods are available, placebo tests can be both statistically more powerful and substantively more credible. 

Formally, a placebo test assesses a conditional independence restriction, which implies that the placebo outcomes are unrelated to treatment assignment once we account for the remaining observed covariates. A limitation of LaLonde’s original test is that it checks only one implication of the full conditional independence assumption: whether the average outcomes, after adjusting for observed covariates, are the same across groups. \citet{imbens2015} discusses other testable implications of this assumption.

\section{Reanalyzing the LaLonde Data}\label{reanalyses}

To demonstrate the methodological advances since \citet{LaLonde} in practice, we revisit the LaLonde data, including the LaLonde-Dehejia-Wahba data, the original LaLonde male samples, and the Lalonde-Cal{\'o}nico-Smith female samples. Our primary focus is on the LaLonde-Dehejia-Wahba data, while results for the other two datasets are reported in the online appendix. For all three datasets, overlap is a central concern. The ATT and CATT are our primary causal estimands.

\subsection{Propensity Scores and Overlap}

We focus on the LaLonde-Dehejia-Wahba (LDW) data because it includes earnings and employment information from 1974. Our analysis uses three LDW datasets: (1) LDW-Experimental, consisting of 185 treated and 280 control individuals from the experimental sample; (2) LDW-CPS, which includes the same treated individuals and 15,922 controls from CPS-SSA-1; and (3) LDW-PSID, comprising the same treated individuals and 2,490 controls from PSID-1. As noted earlier, LaLonde constructed smaller subsamples to improve comparability, but we instead rely on more modern, data-driven methods to assess and address overlap issues.

First, we estimate propensity scores using generalized random forest \citep[GRF,][]{athey2019generalized}, a machine learning method that flexibly models conditional probabilities. Unlike simple logistic regression, GRF can capture complex nonlinearities and higher-order interactions in the covariates, potentially yielding more accurate estimates of the propensity scores. 

\begin{figure}[!th]
    \caption{Assessing Overlap in the Lalonde-Dehejia-Wahba (LDW) Data}\label{fig:overlap}
    \centering
    \begin{minipage}[c]{.3\textwidth}
        \centering
        \captionsetup{justification=centering}
        \begin{subfigure}{\linewidth}
            \includegraphics[width=\linewidth]{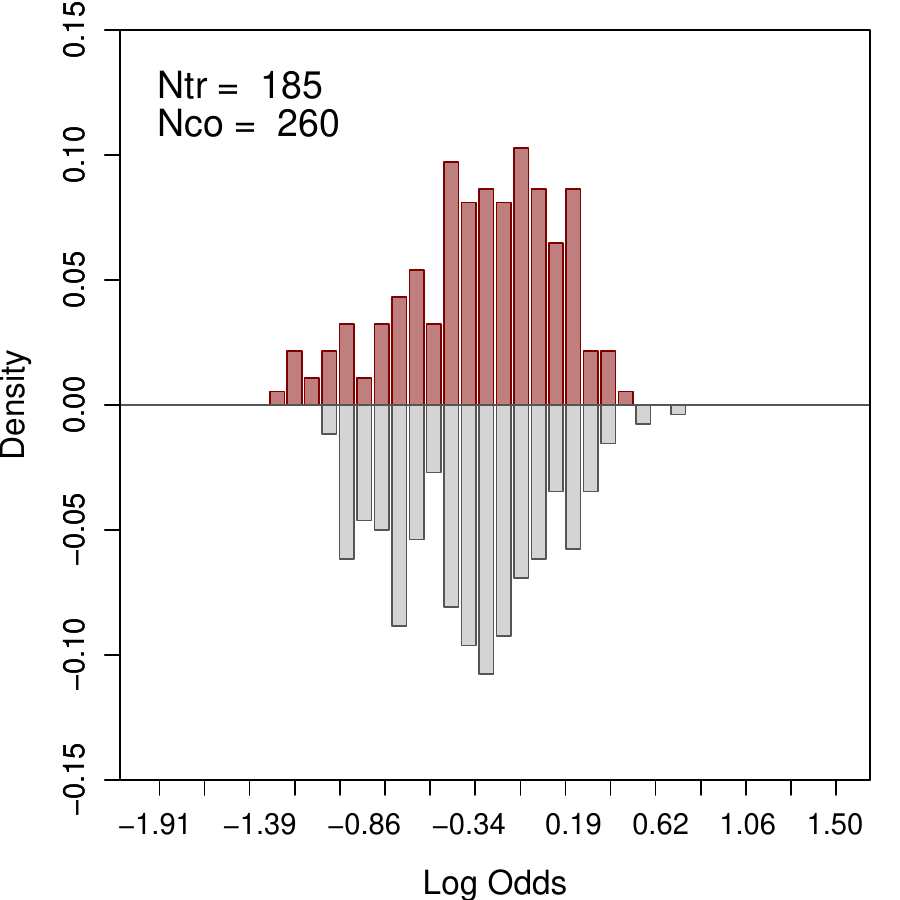}
            \caption{LDW-Experimental}
        \end{subfigure}
    \end{minipage}%
    \begin{minipage}[c]{.65\textwidth}
        \centering
        \captionsetup{justification=centering}
        \begin{subfigure}{0.45\linewidth}
            \includegraphics[width=\linewidth]{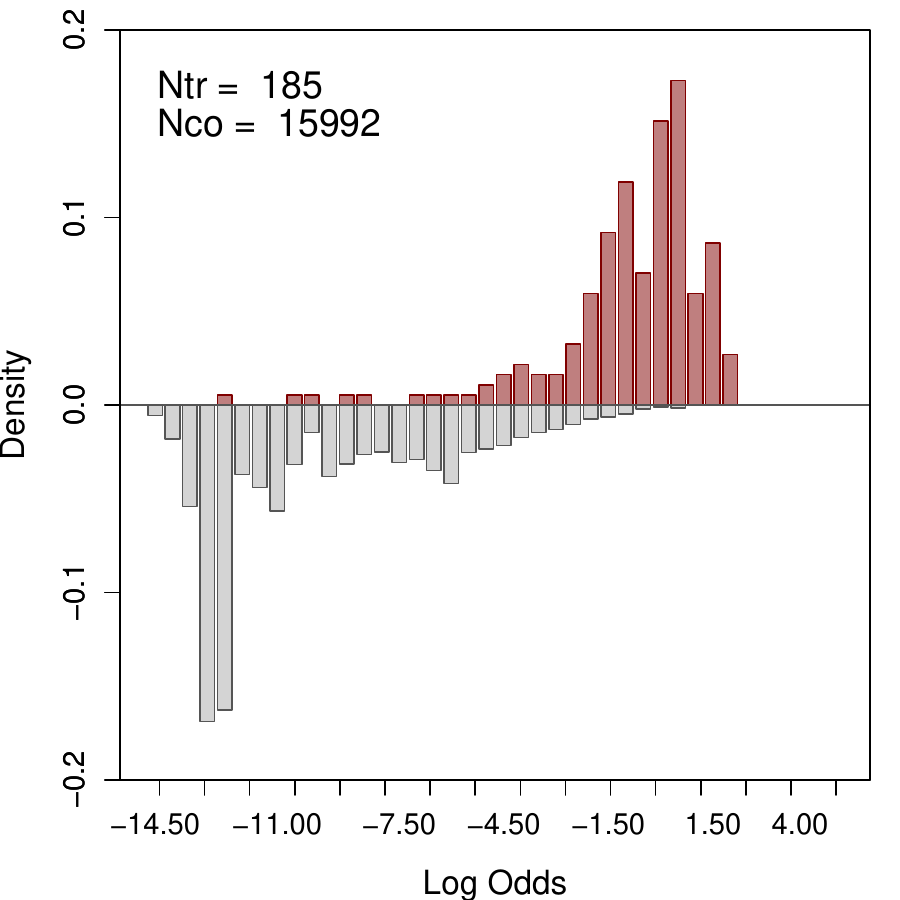}
            \caption{LDW-CPS}
        \end{subfigure}
        \begin{subfigure}{0.45\linewidth}
            \includegraphics[width=\linewidth]{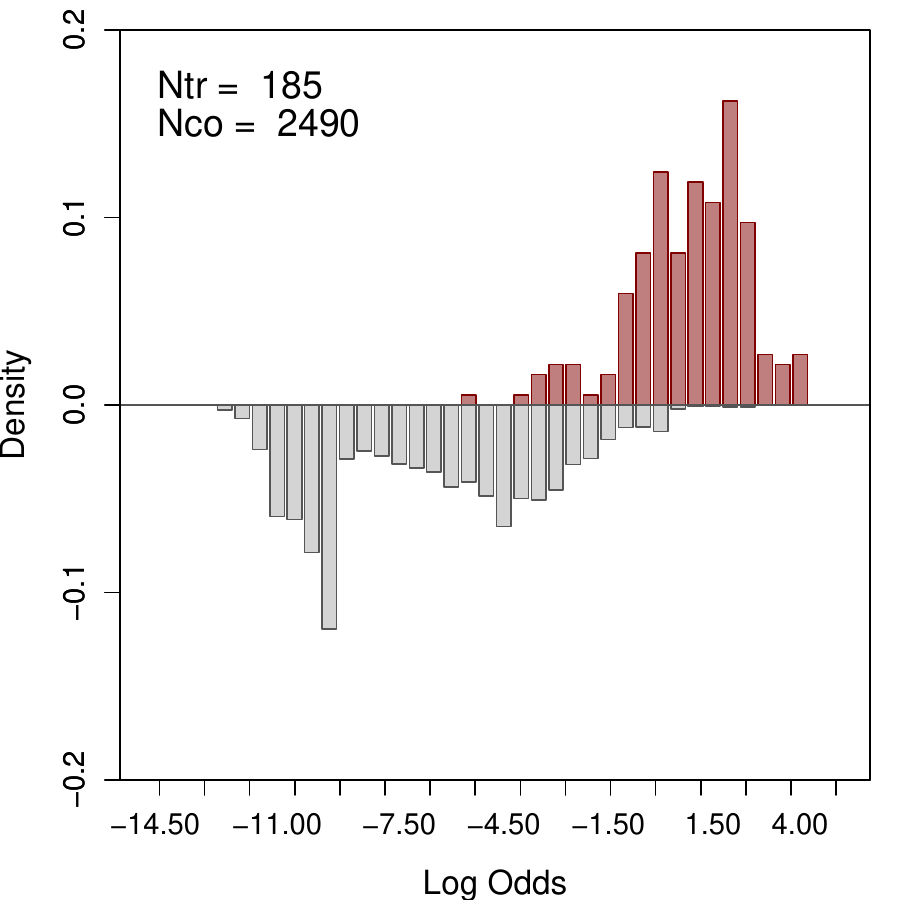}
            \caption{LDW-PSID}
        \end{subfigure}\\
        \begin{subfigure}{0.45\linewidth}
            \includegraphics[width=\linewidth]{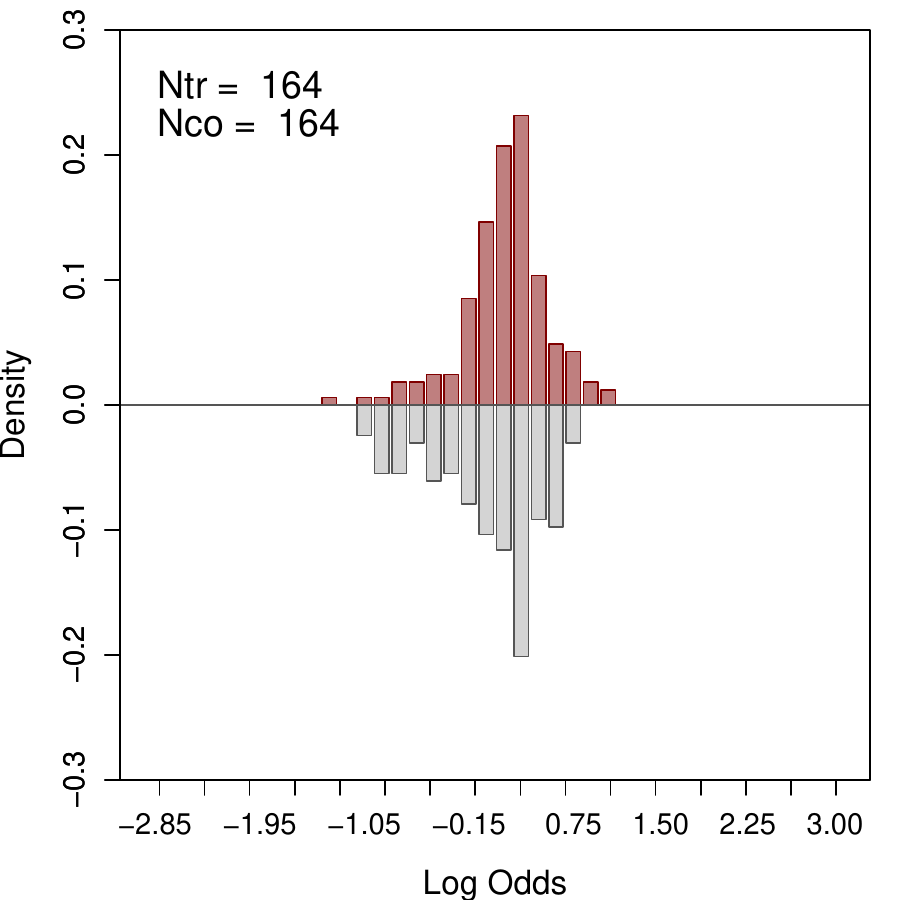}
            \caption{Trimmed LDW-CPS}
        \end{subfigure}
        \begin{subfigure}{0.45\linewidth}
            \includegraphics[width=\linewidth]{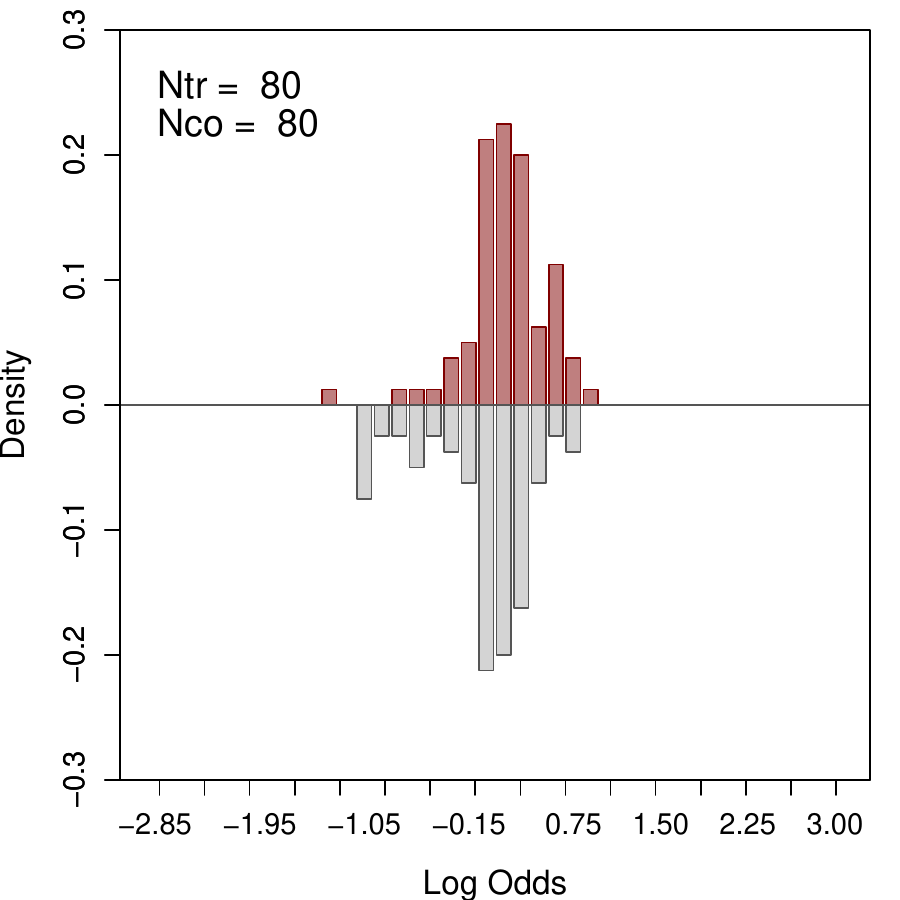}
            \caption{Trimmed LDW-PSID}
        \end{subfigure}
    \end{minipage}%
    \\\raggedright
     {\footnotesize\textbf{\textit{Note:}} Histograms show the log odds ratios for units in the treatment group (top, in maroon) and the control group (bottom, in gray). For unit $i$, the log odds ratio is defined as $\log\frac{\hat{e}}{1 - \hat{e}}$, where $\hat{e}$ is the estimated propensity score from the generalized random forest. Each figure represents a different sample. $N_{tr}$ and $N_{co}$ represent the numbers of treated and control units, respectively. \textbf{A}: LDW-Experimental. \textbf{B}: LDW-CPS. \textbf{C}: LDW-PSID. \textbf{D}: Trimmed LDW-CPS. \textbf{E}: Trimmed LDW-PSID. For D and E, the propensity scores are re-estimated after trimming.}
\end{figure}

Figure~\ref{fig:overlap} (A)–(C) use these estimated scores to assess overlap between treated and control units in each sample. Each panel plots histograms of the log-odds of the estimated propensity scores, defined as $\log\left(\hat{e}/(1-\hat{e})\right)$, where $\hat{e}$ is the estimated propensity score. We use the log-odds scale because it more effectively distinguishes differences at the tails of the distribution. For reference, a log-odds of –3 corresponds roughly to a probability of 0.05, and by symmetry, a log-odds of 3 corresponds to a probability of about 0.95.

In panel (A), the LDW-Experimental sample shows near-perfect overlap: the treated and control groups have closely aligned propensity score distributions. In contrast, panels (B) and (C) reveal severe overlap problems in the nonexperimental samples, with many treated units having propensity scores that fall outside the support of the control group, and large segments of the control group exhibiting extremely low log-odds. Similar patterns are observed in the original LaLonde male samples, as shown in the online appendix.

To address these issues, we construct trimmed versions of the LDW-CPS and LDW-PSID samples. Following the earlier discussion, trimming can improve robustness with only modest loss of precision. We begin by augmenting each nonexperimental sample with the experimental controls and estimating each unit’s probability of being in the experimental data using GRF. We then trim based on preset thresholds, which may exclude some treated units \citep{crump2009dealing}. Next, we re-estimate the propensity scores in the trimmed sample and perform 1:1 matching to refine the control group. This yields two sets of trimmed samples: one with experimental treated units and matched nonexperimental controls, and a second with treated and control units from the experiment, which serves as a benchmark. This two-step trimming and matching procedure improves overlap while preserving a comparison to an experimental benchmark.\footnote{Details of the trimming procedure are provided in the online appendix.} As shown in Figure~\ref{fig:overlap}(D)–(E), overlap improves substantially in both trimmed samples, albeit at the cost of smaller sample sizes.


\subsection{Estimating the ATT} 

We now estimate the ATT using both the original LaLonde-Dehejia-Wahba nonexperimental samples and the newly constructed trimmed samples. We apply a range of estimators, some likely familiar to most readers and others perhaps more novel. Our primary goal is to compare these methods; full computational details are provided in the online appendix.

\begin{figure}[!ht]
    \begin{minipage}[c]{1\textwidth}
    \caption{\\ATT Estimates Given Unconfoundedness: LaLonde-Dehejia-Wahba (LDW) Samples}\label{fig:ldw.att}
    \vspace{-1.5em}\begin{center}
    \includegraphics[height=0.7\linewidth]{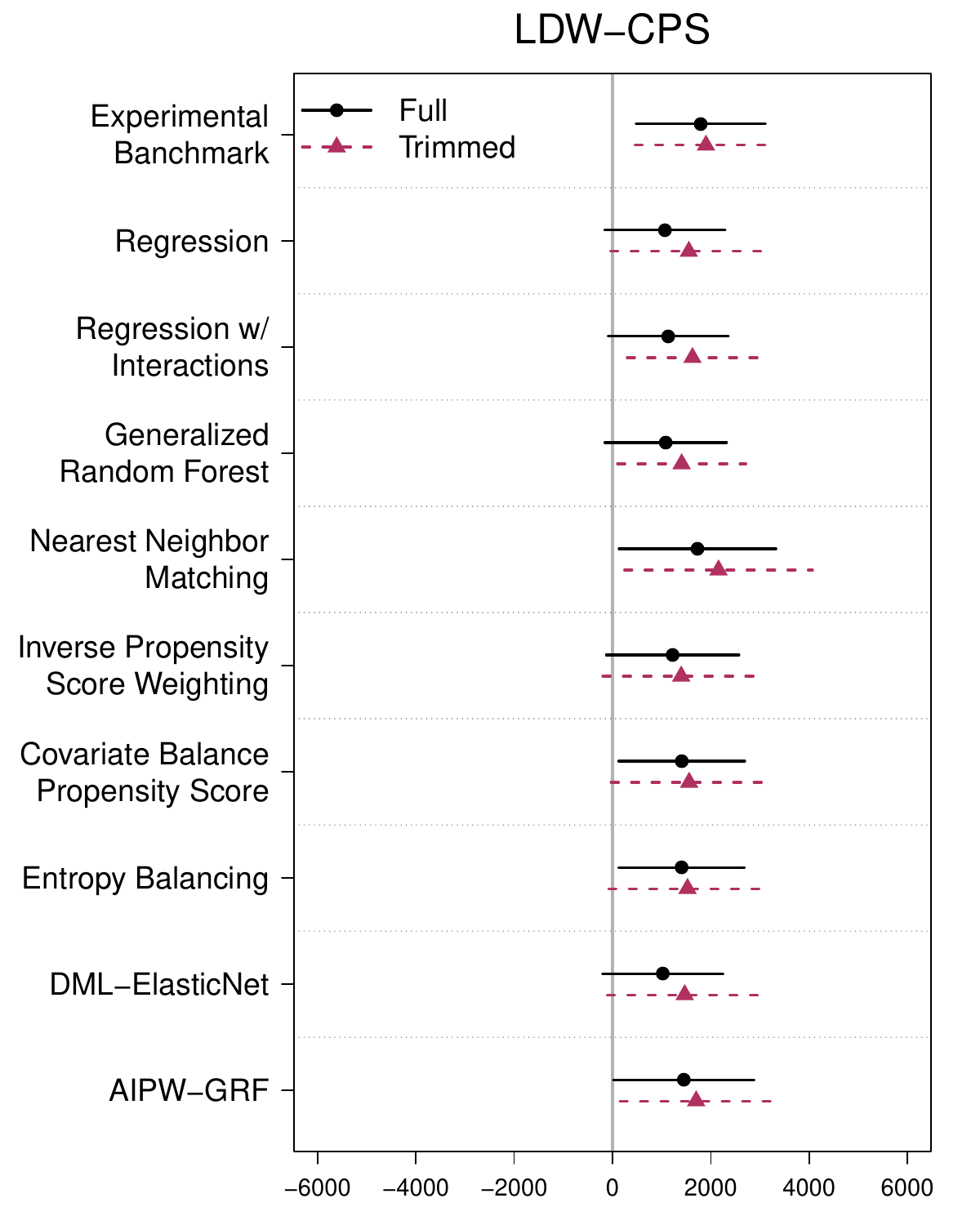}
    \hspace{-1.5em}\includegraphics[height=0.7\linewidth]{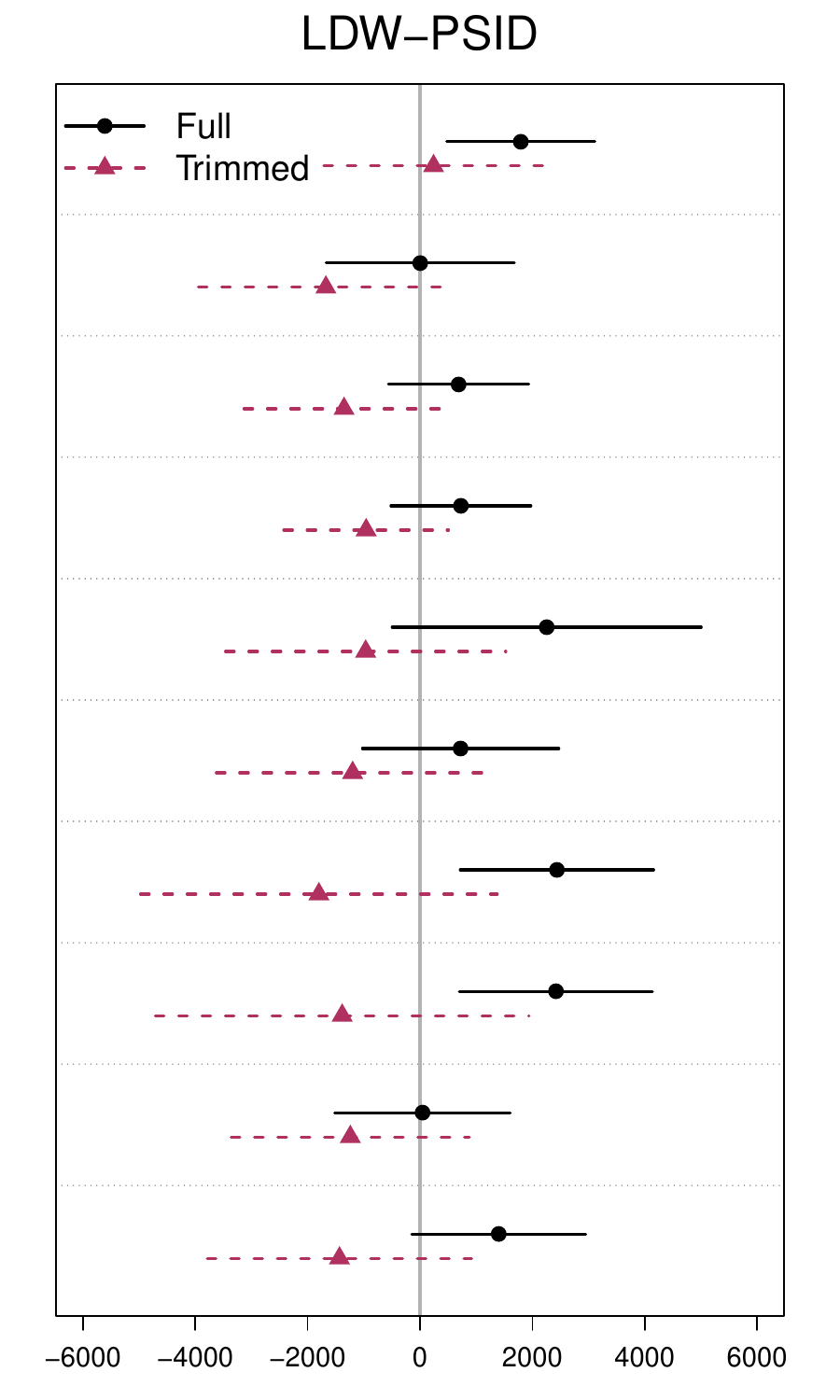}
    \end{center}\vspace{-1em}    
     {\footnotesize\textbf{\textit{Note:}} The figures above show the ATT estimates and their 95\% confidence intervals using four different samples: LDW-CPS and Trimmed LDW-CPS (left panel), and LDW-PSID and Trimmed LDW-PSID (right panel). Estimates based on corresponding experimental samples are presented at the top. Ten estimators are employed, including difference-in-means, linear regression, linear regression with interactions, generalized random forest for outcome modeling, 1:5 nearest neighbor matching with bias correction, inverse propensity score weighting with GRF-estimated propensity scores, covariate-balance propensity score, entropy balancing, double/debiased machine learning with elastic net (DML-ElasticNet), implemented using \texttt{DoubleML}, and augmented inverse propensity score weighting with GRF for both outcome modeling and propensity score estimation (AIPW-GRF), implemented using \texttt{grf}.}
     \end{minipage}
\end{figure}

Figure~\ref{fig:ldw.att} presents the results. The top row reports experimental benchmarks using both the untrimmed and the trimmed LaLonde-Dehejia-Wahba samples. The remaining rows combine the experimental treated units with a nonexperimental control group and apply the following methods: difference-in-means, simple regression; regression with interactions; generalized random forest (GRF) for outcome modeling; nearest neighbor matching with bias correction (matching each treated unit with five control units based on covariates); inverse propensity weighting with GRF-estimated propensity scores; covariate balancing propensity score; entropy balancing; double/debiased machine learning using elastic net (DML-ElasticNet); and augmented inverse propensity weighting via GRF (AIPW-GRF). All estimators use the same ten covariates used in prior analyses.\footnote{The difference-in-means estimates are shown in the online appendix but omitted here due to their extreme values in the LDW-CPS and LDW-PSID samples (\$–8,497 and \$–15,204, respectively). In the trimmed samples, they are closer to the other estimates (\$1,483 and \$–1,505).}

The left panel of Figure~\ref{fig:ldw.att} shows ATT estimates and 95\% confidence intervals from the LDW-CPS sample; the right panel shows results from the LDW-PSID sample. For each method, estimates from the full sample are shown in black, and those from the trimmed data are in red. Solid circles mark point estimates; lines represent 95\% confidence intervals. 

Using the LDW-CPS sample (left panel), all estimators yield positive ATT estimates, though they vary in magnitude. Nearest neighbor matching aligns most closely with the experimental benchmark of \$1,794; covariate balancing propensity score, entropy balancing, and AIPW-GRF also produce estimates near this benchmark. Despite numerical differences, the estimates are not statistically distinguishable from one another. In the trimmed LDW-CPS sample, the estimates show less variation across methods, although confidence intervals widen slightly. All estimates in the trimmed sample center around the experimental benchmark of \$1,911. This stability of estimates after trimming is consistent with the simulation results in \citet{athey2021using}.

Using the LDW-PSID sample (right panel), estimates from the full data show more dispersion, ranging from \$4 to \$2,420. AIPW-GRF yields an estimate closest to the experimental benchmark. In the trimmed LDW-PSID sample, the experimental benchmark is \$306, which is not statistically different from zero at the 5\% level. Although the estimates from the trimmed sample are all negative and more similar to each other, they seem qualitatively different from the experimental benchmark. However, due to large standard errors, these differences are not statistically significant.  

Overall, the results suggest that improving overlap based on observed covariates reduces model dependence and estimate variability, yielding more robust estimates of the statistical estimand. However, the fact that many methods produce ATT estimates close to the experimental benchmark using LDW-CPS may have given researchers a false sense of confidence that modern estimators can recover causal effects---even in settings where the unconfoundedness assumption is likely violated.

\subsection{Treatment Effect Heterogeneity.}

We explore treatment effect heterogeneity by estimating the CATT, comparing estimates from the experimental and nonexperimental samples in the LaLonde-Dehejia-Wahba data. For simplicity, we focus on the LDW-CPS data—both full and trimmed—for the nonexperimental analyses, using the corresponding experimental samples for benchmarking. In the trimmed sample, 21 treated units lacking comparable controls are removed, and the nonexperimental control group is trimmed using one-to-one matching on the estimated propensity score with the remaining 164 treated units. Details on the trimming procedure are discussed earlier in the paper and in the online appendix. We estimate CATT using a causal forest estimator \citep{athey2017generalized}, omitting technical details.

\begin{figure}[!ht]
    \caption{CATT Estimates: Experimental versus Non-Experimental}\label{fig:catt}
    \begin{minipage}[c]{1\textwidth}
    \begin{subfigure}{0.47\linewidth}
        \includegraphics[width=1\linewidth]{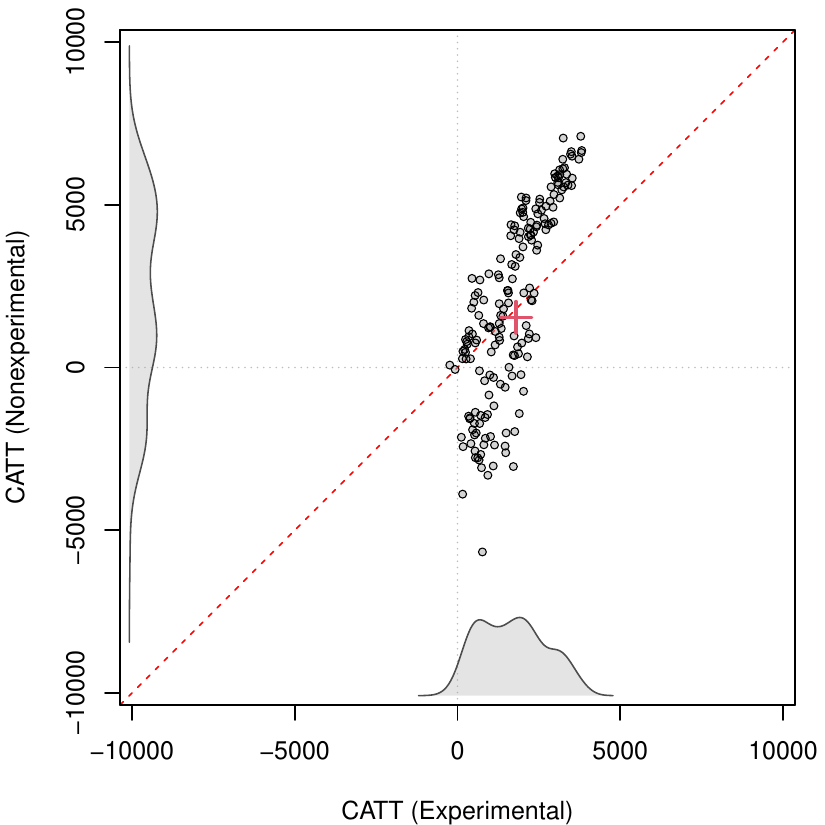}
        \caption{\centering LDW-CPS}            
    \end{subfigure}\hspace{1em}
    \begin{subfigure}{0.47\linewidth}
        \includegraphics[width=1\linewidth]{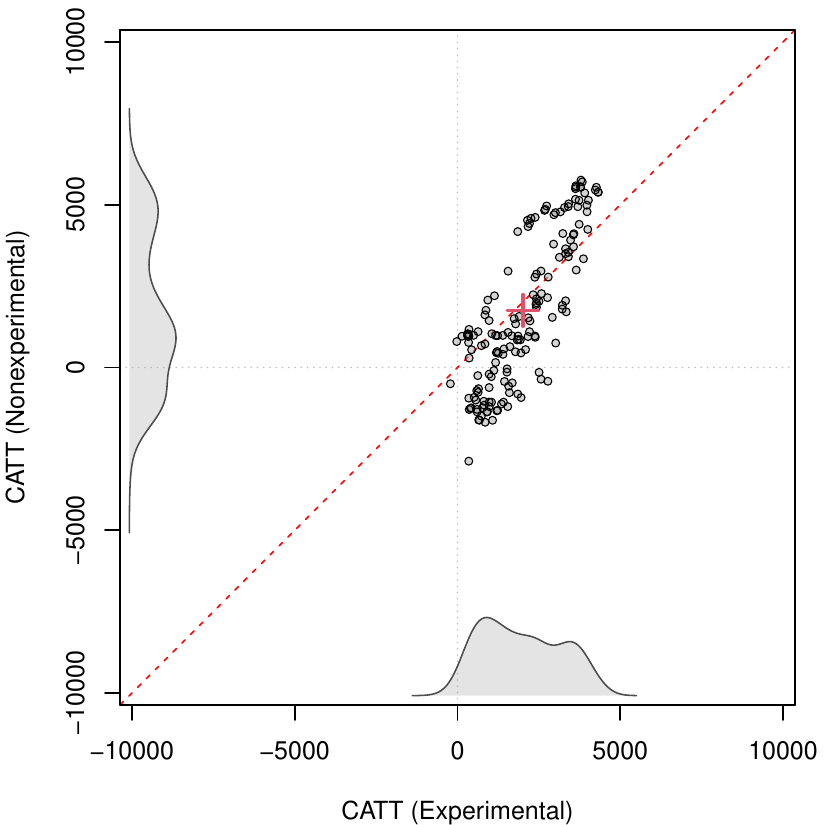}
        \caption{\centering Trimmed LDW-CPS}
    \end{subfigure}
    \end{minipage}
    \\\raggedright
     {\footnotesize\textbf{\textit{Note:}} In the above figures, the scatterplots compare CATT estimates from the experimental data (x-axis) and the nonexperimental data (y-axis). \textbf{Subfigure A} compares CATT estimates from LDW-Experimental vs. LDW-CPS, while \textbf{Subfigure B} compares those from their trimmed versions, which exclude individuals with extreme estimated propensity scores. In each figure, a dot represents a CATT estimate based on the covariate values of a treated unit; the red cross marks the pair of ATT estimates from the two samples. The marginal distributions of the CATT estimates are shown as density plots along the axes—the bottom plot for the experimental data and the left-side plot for the nonexperimental data. The red dashed line indicates the 45-degree line. Uncertainty estimates are not shown in this figure.}
\end{figure}

Figure~\ref{fig:catt} plots the estimated CATT at the covariate values of each treated unit, with experimental data on the x-axis and nonexperimental data on the y-axis, using both the full sample and the trimmed sample. Each dot represents a pair of CATT estimates for an individual with a given covariate profile. The red cross marks the pair of ATT estimates obtained using the AIPW-GRF estimator. 

Figure~\ref{fig:catt} reveals several interesting patterns. First, there is substantial heterogeneity in the estimated treatment effects: CATT estimates based on the experimental data range from \$–236 to \$3,817 in the full sample and from \$–218 to \$4,324 in the trimmed sample. Second, in the full sample, although the AIPW-GRF estimator yields ATT estimates that closely match the experimental benchmark, the CATT estimates from the experimental and nonexperimental data diverge sharply from the 45-degree line. In particular, the nonexperimental CATT estimates span a much wider range—from \$–5,667 to \$7,102—far exceeding that of the experimental estimates, with over one-fourth of treated units obtaining negative CATT estimates. Third, improving overlap substantially enhances the robustness of the CATT estimates: in the trimmed sample, CATT estimates from the experimental and nonexperimental data roughly align along the 45-degree line, and the range of the nonexperimental estimates—\$–2,877 to \$5,758—is much narrower than in the untrimmed sample.

Applying the same procedure to the LDW-PSID data yields CATT estimates ranging from \$–8,422 to \$4,870 in the full sample and from \$–5,088 to \$1,571 in the trimmed sample—showing greater deviation from the experimental benchmark than in the LDW-CPS case. Analysis of the LaLonde male sample and the reconstructed female data further demonstrates that recovering the CATT is substantially more difficult than estimating the ATT. The combination of the experimental subsample constructed by \citet{dehejiawahba} and the CPS nonexperimental comparison group appears to be an outlier in its ability to recover experimental benchmarks. These additional results are reported in the online appendix.

\subsection{Validation Through Placebo Analyses} 

While modern nonexperimental methods may be effective in estimating the statistical estimand---the covariate-adjusted difference in average outcomes between treated and control groups---this does {\it not} imply that the estimate approximates the causal estimand, such as the ATT. Identifying the ATT requires the unconfoundedness assumption, which is fundamentally untestable. However, we can assess its plausibility through placebo analyses.

In the LaLonde setting, we use 1975 earnings as a placebo outcome and exclude both 1975 earnings and employment status from the set of conditioning variables. By construction, 1975 earnings could not have been affected by the treatment, which occurred afterward. We also construct two new trimmed samples, omitting these variables during the trimming process. We then estimate the ATT for the placebo outcome, adjusting for the remaining covariates using a range of estimators.

If the ATT estimates for the placebo outcome are close to zero, this lends support to the unconfoundedness assumption, as it suggests that even without conditioning on 1975 earnings or employment status, treatment assignment is likely independent of the untreated potential outcome $Y_i(0)$—that is, 1978 earnings had the individual not participated in the program. Including these variables in the main analysis would then make unconfoundedness even more credible. However, if the placebo ATT estimates differ significantly from zero, this may indicate either that 1975 earnings and employment status are key confounders, or that unmeasured factors such as perseverance influence both treatment assignment and outcomes. In that case, the placebo analysis fails to bolster unconfoundedness in the main analysis.

\begin{figure}[!ht]    
    \begin{minipage}[c]{1\textwidth}
    \caption{\\Placebo Tests: 1975 Earnings as the Outcome}\label{fig:ldw.placebo}\vspace{-1.5em}
    \begin{center}
    \includegraphics[height=0.7\linewidth]{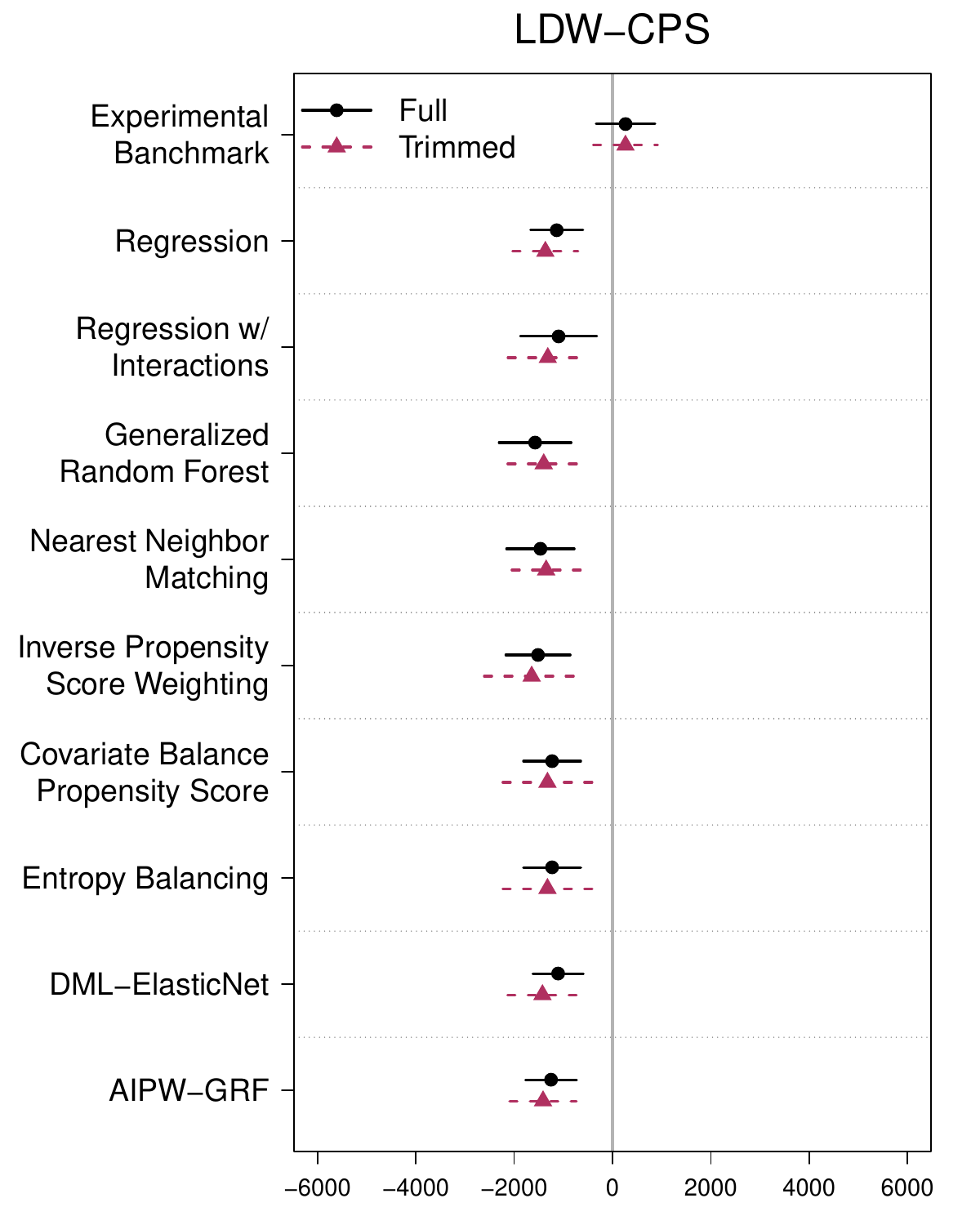}
    \hspace{-1.5em}\includegraphics[height=0.7\linewidth]{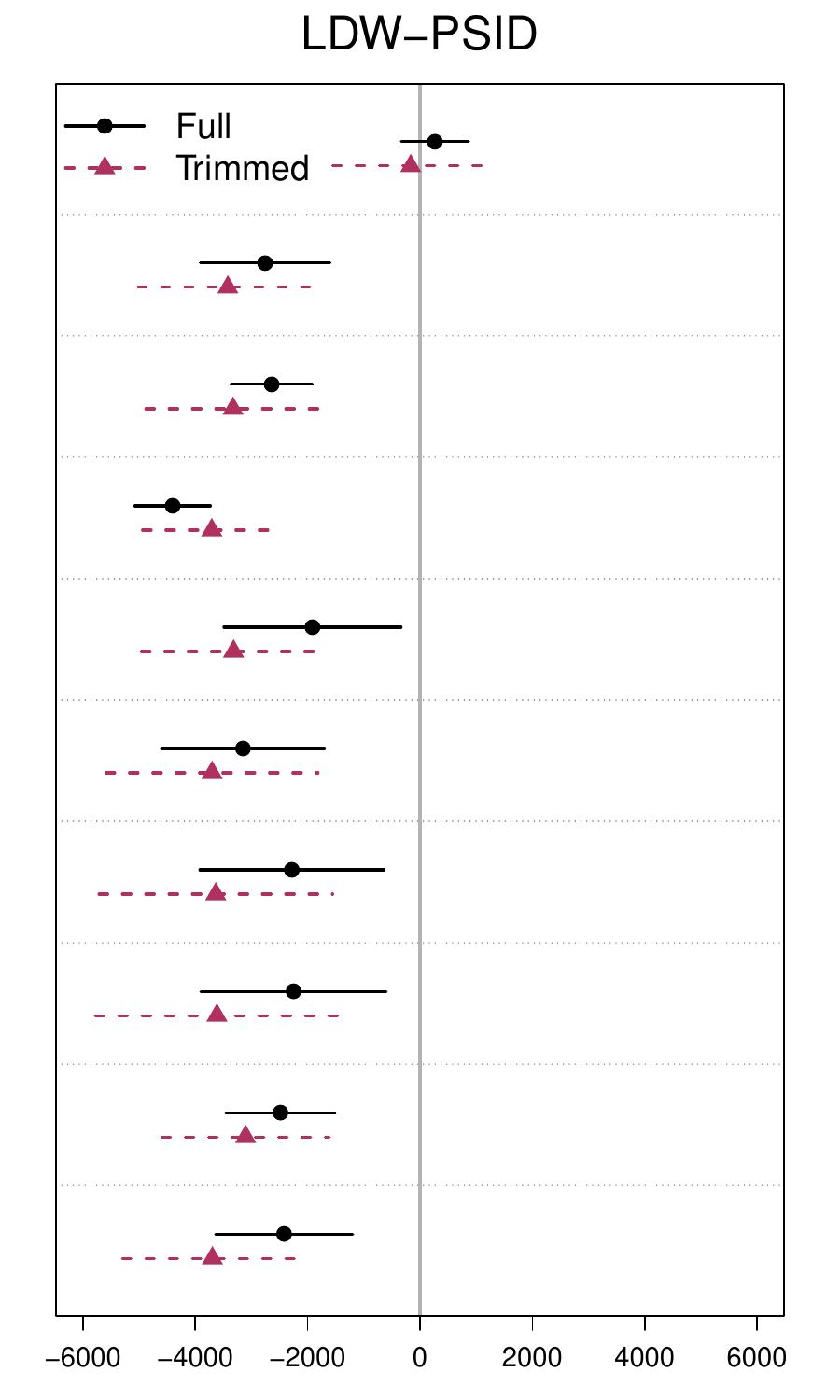}
    \end{center}  \vspace{-1em}  
     {\footnotesize\textbf{\textit{Note:}} The figures above show the placebo estimates and their 95\% confidence intervals using four samples: LDW-CPS and Trimmed LDW-CPS (left panel), and LDW-PSID and Trimmed LDW-PSID (right panel). For each case, results based on the corresponding experimental samples are shown at the top as benchmarks. All estimates are produced using the same ten estimators introduced earlier.}
     \end{minipage}
\end{figure}

Figure~\ref{fig:ldw.placebo} presents the results. As expected, the experimental benchmarks are close to zero and statistically insignificant. In contrast, all estimators using nonexperimental data yield large, negative estimates. While trimming improves the stability of these estimates, they remain statistically different from zero.\footnote{The difference-in-means estimates are not shown here; they are \$–12,118, \$–17,531, \$–14,056, and \$–4,670 across the four panels, respectively. Details are available in the online appendix.} Moreover, as shown in the online appendix, CATT estimates for a similar placebo test based on experimental data cluster around zero, while their nonexperimental counterparts are consistently negative and sizable, indicating substantial bias. These patterns further illustrates the point we aim to emphasize:  With sufficient overlap, modern methods can robustly estimate the statistical estimands, but not necessarily the causal estimands if unconfoundedness is violated.

\subsection{Alternative Samples} 

For comparison, we also revisit the original male samples used in \citet{LaLonde}. These datasets lack information on 1974 earnings and employment status. We find that with sufficient overlap, most modern estimators yield estimates within relatively narrow ranges when using CPS-SSA-1 or PSID-1 as control groups. However, these estimates—most of which are negative—do not align with the experimental ATT benchmarks. Similar patterns are reported by \citet{smith2001reconciling, smith2005does}. Using these nonexperimental data, modern methods also fail to recover the experimental CATT benchmarks, as noted earlier.

Using the LaLonde-Cal{\'o}nico-Smith female sample with PSID data as nonexperimental controls, we find that many modern methods produce estimates close to the experimental benchmarks, though standard errors are often quite large. As noted by \citet{calonico2017women}, selection appears to be less severe for female participants than for male participants in the National Supported Work program. However, overlap remains a significant challenge. Moreover, a placebo test using the number of children in 1975—a variable not included in LaLonde’s original analysis—does not support the unconfoundedness assumption. CATT estimates also fail to recover the experimental benchmarks.

As an example of a dataset that readily passes a placebo test, we include in the appendix a reanalysis of data from \citet{imbensrubinsacerdote}, who conducted an original survey to study the impact of lottery prize size in Massachusetts during the mid-1980s on the economic behavior of lottery players. The primary outcome is post-winning labor earnings, with data available for earnings in the six preceding years. While lottery outcomes may seem random, there are systematic differences in pre-treatment variables, likely related to the number of tickets purchased or survey response rates. However, in this case, placebo tests provide strong evidence supporting the unconfoundedness assumption, bolstering the credibility of the causal estimates. The availability of six pretreatment earnings measures proves particularly valuable: they likely capture both selection and outcome-relevant factors and serve as strong placebo outcomes due to their comparability to the post-treatment outcome. We report the details in the online appendix.

\subsection{Summary}

Our reexamination of the LaLonde data shows that when overlap is ensured, the choice of estimation method becomes less consequential, as most methods yield similar results. However, these estimates may not represent the causal quantity of interest if the unconfoundedness assumption is violated. Supplementary analyses, such as placebo tests, can help assess how plausible this assumption is. Although in both the LDW-CPS and LaLonde-Cal{\'o}nico-Smith datasets, many modern methods appear to recover the experimental benchmark for the ATT, placebo tests fail to support  unconfoundedness. In typical research settings where experimental benchmarks are unavailable, we cannot know whether such estimates credibly recover the causal estimand. Indeed, with the original LaLonde male sample or the LDW-PSID dataset, modern estimators fail to recover the ATT. 

The answer to whether modern nonexperimental methods can credibly estimate causal effects, forty years after LaLonde's critique, is therefore nuanced. Researchers can now use data-driven approaches to improve overlap and apply modern methods to reliably estimate the statistical estimand. However, credible causal inference still depends on the validity of the unconfoundedness assumption. If supplementary analyses—such as placebo tests—support this assumption, modern methods can yield highly credible estimates. Otherwise, they cannot.


\section{Lessons Learned}

What lessons has the methodological literature since \citet{LaLonde} taught us? And what specific analyses should researchers conduct today in similar nonexperimental studies? We return to the five lessons and offer practical recommendations.

First, any analysis of causal effects using nonexperimental data should begin with a careful investigation of the treatment assignment mechanism. A clear understanding of the ``design'' is essential for evaluating the plausibility of the unconfoundedness assumption. The case for relying on unconfoundedness is strongest when researchers believe that selection into treatment is driven by factors that are well understood, observed, and measured. When that condition holds, flexibly adjusting for observed pretreatment covariates can reduce reliance on strong modeling assumptions. In cases where important confounders are unobserved, researchers may turn to panel data methods that account for time-invariant unobserved confounding; for recent overviews, see \citet{xu2023causal} and \citet{arkhangelsky2023causal}.

Second, the literature has underscored the importance of assessing and improving overlap in covariate distributions. The comparison groups LaLonde used differed substantially from the experimental sample, prompting him to discard some units based on age, employment status, and earnings. Since then, more systematic and data-driven approaches have been developed to diagnose and address lack of overlap, including propensity score-based trimming and weighting strategies. The loss of efficiency is often modest and a worthwhile cost.

Third, relatedly, the propensity score has become central to both diagnosing overlap and estimating treatment effects. Researchers now routinely estimate the propensity score using flexible methods and evaluate overlap by comparing the distribution of propensity scores across treated and control groups. When necessary, samples can be trimmed to improve comparability. While the concept of the propensity score was introduced just prior to LaLonde’s work in \citet{rosenbaum1983central}, it has since become foundational. Doubly robust estimators, particularly those that combine outcome modeling with inverse propensity score weighting, yield consistent estimates when either the outcome model or the propensity score model is correctly specified. The integration of machine learning techniques into causal inference has further strengthened these methods by reducing dependence on \emph{ad hoc} specification choices. We expect these methods to see broader adoption among economists and social scientists.

Fourth, attention has shifted from solely estimating average treatment effects to understanding treatment effect heterogeneity. Policymakers increasingly ask not only ``Does it work?'' but ``For whom does it work?'' and ``Where does it cause harm?'' The literature has developed tools to estimate conditional average treatment effects and quantile treatment effects. These estimands help unpack heterogeneous impacts and can inform personalized policy decisions. Large datasets and algorithmic advances have made it easier to estimate these effects flexibly and at scale \citep[{\it e.g.},][]{wager2018estimation}. We encourage the estimation and visualization of these additional quantities of interest.

Finally, the credibility of causal estimates increasingly hinges on validation exercises—especially placebo tests. While LaLonde included some placebo analyses using lagged earnings, his main focus was on comparing nonexperimental estimates to experimental benchmarks. In contrast, modern practice places greater emphasis on formal diagnostic checks. Placebo tests, such as estimating effects on outcomes known to be unaffected by the treatment, offer informative checks on key identifying assumptions like unconfoundedness and should be more routinely used in empirical analyses.

To help researchers apply these lessons in practice, we provide a detailed online tutorial—including \texttt{R} code and the data used in this paper and the online appendix—available at \url{https://yiqingxu.org/tutorials/lalonde/}. The tutorial replicates our analyses and can be easily adapted to other datasets.

\vspace{3em}\noindent {\it\small $\blacksquare\quad$ We thank the Office of Naval Research for support under grant numbers N00014-17-1-2131 and N00014-19-1-2468, and Amazon for a gift. We are grateful to Susan Athey, Scott Cunningham, Alexis Diamond, Peng Ding, Dean Eckles, and Xiang Zhou for helpful comments and feedback, and to Timothy Taylor, Jonathan Parker, and Heidi Williams for detailed editorial suggestions. We also thank Zihan Xie and Jinwen Wu for excellent research assistance.}

\vspace{2em}
\small
\bibliography{\bib}

\begin{thebibliography}{78}
\providecommand{\natexlab}[1]{#1}
\providecommand{\url}[1]{\texttt{#1}}
\expandafter\ifx\csname urlstyle\endcsname\relax
  \providecommand{\doi}[1]{doi: #1}\else
  \providecommand{\doi}{doi: \begingroup \urlstyle{rm}\Url}\fi

\bibitem[Abadie and Cattaneo(2018)]{abadie2018econometric}
Alberto Abadie and Matias~D Cattaneo.
\newblock Econometric methods for program evaluation.
\newblock \emph{Annual Review of Economics}, 10:\penalty0 465--503, 2018.

\bibitem[Abadie and Imbens(2006)]{abadie2006}
Alberto Abadie and Guido~W Imbens.
\newblock Large sample properties of matching estimators for average treatment effects.
\newblock \emph{Econometrica}, 74\penalty0 (1):\penalty0 235--267, 2006.

\bibitem[Abadie and Imbens(2008)]{abadie2008failure}
Alberto Abadie and Guido~W Imbens.
\newblock On the failure of the bootstrap for matching estimators.
\newblock \emph{Econometrica}, 76\penalty0 (6):\penalty0 1537--1557, 2008.

\bibitem[Abadie and Imbens(2011)]{abadie2011bias}
Alberto Abadie and Guido~W Imbens.
\newblock Bias-corrected matching estimators for average treatment effects.
\newblock \emph{Journal of Business \& Economic Statistics}, 29\penalty0 (1):\penalty0 1--11, 2011.

\bibitem[Abadie and Imbens(2016)]{abadie2016matching}
Alberto Abadie and Guido~W Imbens.
\newblock Matching on the estimated propensity score.
\newblock \emph{Econometrica}, 84\penalty0 (2):\penalty0 781--807, 2016.

\bibitem[Angrist and Pischke(2008)]{angrist2008mostly}
Joshua~D Angrist and J{\"o}rn-Steffen Pischke.
\newblock \emph{Mostly harmless econometrics: An empiricist's companion}.
\newblock Princeton University Press, 2008.

\bibitem[Angrist and Pischke(2010)]{angrist2010}
Joshua~D Angrist and Jorn-Steffen Pischke.
\newblock The credibility revolution in empirical economics: How better research design is taking the con out of econometrics.
\newblock \emph{Journal of Economic Perspectives}, 24\penalty0 (2):\penalty0 3--30, 2010.

\bibitem[Arkhangelsky and Imbens(2024)]{arkhangelsky2023causal}
Dmitry Arkhangelsky and Guido Imbens.
\newblock Causal models for longitudinal and panel data: A survey.
\newblock \emph{The Econometrics Journal}, 2024.

\bibitem[Ashenfelter(1978)]{ashenfelter1978estimating}
Orley Ashenfelter.
\newblock Estimating the effect of training programs on earnings.
\newblock \emph{The Review of Economics and Statistics}, pages 47--57, 1978.

\bibitem[Ashenfelter and Card(1985)]{ashenfelter1985using}
Orley Ashenfelter and David Card.
\newblock Using the longitudinal structure of earnings to estimate the effect of training programs.
\newblock \emph{The Review of Economics and Statistics}, 67\penalty0 (4):\penalty0 648--660, 1985.

\bibitem[Athey and Imbens(2016)]{athey2016recursive}
Susan Athey and Guido Imbens.
\newblock Recursive partitioning for heterogeneous causal effects.
\newblock \emph{Proceedings of the National Academy of Sciences}, 113\penalty0 (27):\penalty0 7353--7360, 2016.

\bibitem[Athey and Imbens(2017)]{athey2017state}
Susan Athey and Guido~W Imbens.
\newblock The state of applied econometrics: Causality and policy evaluation.
\newblock \emph{Journal of Economic Perspectives}, 31\penalty0 (2):\penalty0 3--32, 2017.

\bibitem[Athey and Imbens(2019)]{athey2019machine}
Susan Athey and Guido~W Imbens.
\newblock Machine learning methods that economists should know about.
\newblock \emph{Annual Review of Economics}, 11:\penalty0 685--725, 2019.

\bibitem[Athey et~al.(2017)Athey, Tibshirani, and Wager]{athey2017generalized}
Susan Athey, Julie Tibshirani, and Stefan Wager.
\newblock Generalized random forests.
\newblock \emph{arXiv preprint arXiv:1610.01271}, 2017.
\newblock URL \url{https://arxiv.org/abs/1610.01271}.

\bibitem[Athey et~al.(2018)Athey, Imbens, and Wager]{athey2018approximate}
Susan Athey, Guido~W Imbens, and Stefan Wager.
\newblock Approximate residual balancing: debiased inference of average treatment effects in high dimensions.
\newblock \emph{Journal of the Royal Statistical Society: Series B (Statistical Methodology)}, 80\penalty0 (4):\penalty0 597--623, 2018.

\bibitem[Athey et~al.(2019)Athey, Tibshirani, Wager, et~al.]{athey2019generalized}
Susan Athey, Julie Tibshirani, Stefan Wager, et~al.
\newblock Generalized random forests.
\newblock \emph{The Annals of Statistics}, 47\penalty0 (2):\penalty0 1148--1178, 2019.

\bibitem[Athey et~al.(2021)Athey, Imbens, Metzger, and Munro]{athey2021using}
Susan Athey, Guido~W Imbens, Jonas Metzger, and Evan Munro.
\newblock Using wasserstein generative adversarial networks for the design of monte carlo simulations.
\newblock \emph{Journal of Econometrics}, 2021.

\bibitem[Bang and Robins(2005)]{bang2005doubly}
Heejung Bang and James~M Robins.
\newblock Doubly robust estimation in missing data and causal inference models.
\newblock \emph{Biometrics}, 61\penalty0 (4):\penalty0 962--973, 2005.

\bibitem[Barnow et~al.(1980)Barnow, Cain, Goldberger, et~al.]{barnow1980issues}
Burt~S Barnow, Glen~George Cain, Arthur~Stanley Goldberger, et~al.
\newblock \emph{Issues in the analysis of selectivity bias}.
\newblock University of Wisconsin, Inst. for Research on Poverty, 1980.

\bibitem[Cal{\'o}nico and Smith(2017)]{calonico2017women}
Sebastian Cal{\'o}nico and Jeffrey Smith.
\newblock The women of the national supported work demonstration.
\newblock \emph{Journal of Labor Economics}, 35\penalty0 (S1):\penalty0 S65--S97, 2017.

\bibitem[Chernozhukov et~al.(2017)Chernozhukov, Chetverikov, Demirer, Duflo, Hansen, and Newey]{chernozhukov2017double}
Victor Chernozhukov, Denis Chetverikov, Mert Demirer, Esther Duflo, Christian Hansen, and Whitney Newey.
\newblock Double/debiased/neyman machine learning of treatment effects.
\newblock \emph{American Economic Review, Papers and Proceedings}, 107\penalty0 (5):\penalty0 261--65, 2017.

\bibitem[Chernozhukov et~al.(2018)Chernozhukov, Chetverikov, Demirer, Duflo, Hansen, Newey, and Robins]{chernozhukov2018double}
Victor Chernozhukov, Denis Chetverikov, Mert Demirer, Esther Duflo, Christian Hansen, Whitney Newey, and James Robins.
\newblock Double/debiased machine learning for treatment and structural parameters.
\newblock \emph{The Econometrics Journal}, 21\penalty0 (1), 2018.

\bibitem[Chernozhukov et~al.(2024)Chernozhukov, Hansen, Kallus, Spindler, and Syrgkanis]{chernozhukov2024applied}
Victor Chernozhukov, Christian Hansen, Nathan Kallus, Martin Spindler, and Vasilis Syrgkanis.
\newblock Applied causal inference powered by ml and ai.
\newblock \emph{arXiv preprint arXiv:2403.02467}, 2024.

\bibitem[Cinelli et~al.(2022)Cinelli, Forney, and Pearl]{cinelli2022crash}
Carlos Cinelli, Andrew Forney, and Judea Pearl.
\newblock A crash course in good and bad controls.
\newblock \emph{Sociological Methods \& Research}, page 00491241221099552, 2022.

\bibitem[Corporation(1980)]{mdrc1980}
Manpower Development~Research Corporation.
\newblock Summary and findings of the national subsidized work demonstration, 1980.
\newblock URL \url{https://www.mdrc.org/sites/default/files/full_249.pdf}.

\bibitem[Crump et~al.(2006)Crump, Hotz, Imbens, and Mitnik]{crump2006moving}
Richard~K Crump, V~Joseph Hotz, Guido Imbens, and Oscar Mitnik.
\newblock Moving the goalposts: Addressing limited overlap in the estimation of average treatment effects by changing the estimand, 2006.

\bibitem[Crump et~al.(2009)Crump, Hotz, Imbens, and Mitnik]{crump2009dealing}
Richard~K Crump, V~Joseph Hotz, Guido~W Imbens, and Oscar~A Mitnik.
\newblock Dealing with limited overlap in estimation of average treatment effects.
\newblock \emph{Biometrika}, pages 187--199, 2009.

\bibitem[Cunningham(2018)]{cunningham2018causal}
Scott Cunningham.
\newblock \emph{Causal inference: The mixtape}.
\newblock Yale University Press, 2018.

\bibitem[Currie et~al.(2020)Currie, Kleven, and Zwiers]{currie2020technology}
Janet Currie, Henrik Kleven, and Esm{\'e}e Zwiers.
\newblock Technology and big data are changing economics: Mining text to track methods.
\newblock In \emph{AEA Papers and Proceedings}, volume 110, pages 42--48, 2020.

\bibitem[Dehejia and Wahba(1999)]{dehejiawahba}
Rajeev~H Dehejia and Sadek Wahba.
\newblock Causal effects in nonexperimental studies: Reevaluating the evaluation of training programs.
\newblock \emph{Journal of the American statistical Association}, 94\penalty0 (448):\penalty0 1053--1062, 1999.

\bibitem[Dehejia and Wahba(2002)]{dehejia2002propensity}
Rajeev~H Dehejia and Sadek Wahba.
\newblock Propensity score-matching methods for nonexperimental causal studies.
\newblock \emph{Review of Economics and statistics}, 84\penalty0 (1):\penalty0 151--161, 2002.

\bibitem[Diamond and Sekhon(2013)]{diamond2013genetic}
Alexis Diamond and Jasjeet~S Sekhon.
\newblock Genetic matching for estimating causal effects: A general multivariate matching method for achieving balance in observational studies.
\newblock \emph{Review of Economics and Statistics}, 95\penalty0 (3):\penalty0 932--945, 2013.

\bibitem[Ding(2024)]{ding2024first}
Peng Ding.
\newblock \emph{A first course in causal inference}.
\newblock CRC Press, 2024.

\bibitem[Firpo(2007)]{firpo2007efficient}
Sergio Firpo.
\newblock Efficient semiparametric estimation of quantile treatment effects.
\newblock \emph{Econometrica}, 75\penalty0 (1):\penalty0 259--276, 2007.

\bibitem[Fraker and Maynard(1987)]{fraker1987adequacy}
Thomas Fraker and Rebecca Maynard.
\newblock The adequacy of comparison group designs for evaluations of employment-related programs.
\newblock \emph{Journal of Human Resources}, pages 194--227, 1987.

\bibitem[Hainmueller(2012)]{hainmueller}
Jens Hainmueller.
\newblock Entropy balancing for causal effects: A multivariate reweighting method to produce balanced samples in observational studies.
\newblock \emph{Political Analysis}, 20\penalty0 (1):\penalty0 25--46, 2012.

\bibitem[Heckman(1978)]{heckman1978}
James~J. Heckman.
\newblock Dummy endogenous variables in a simultaneous equation system.
\newblock \emph{Econometrica}, 46\penalty0 (4):\penalty0 931--959, 1978.

\bibitem[Heckman and Hotz(1989)]{heckman1989choosing}
James~J Heckman and V~Joseph Hotz.
\newblock Choosing among alternative nonexperimental methods for estimating the impact of social programs: The case of manpower training.
\newblock \emph{Journal of the American statistical Association}, 84\penalty0 (408):\penalty0 862--874, 1989.

\bibitem[Heckman et~al.(1987)Heckman, Hotz, and Dabos]{heckman1987we}
James~J Heckman, V~Joseph Hotz, and Marcelo Dabos.
\newblock Do we need experimental data to evaluate the impact of manpower training on earnings?
\newblock \emph{Evaluation Review}, 11\penalty0 (4):\penalty0 395--427, 1987.

\bibitem[Heckman et~al.(1997)Heckman, Ichimura, and Todd]{heckman1997matching}
James~J Heckman, Hidehiko Ichimura, and Petra~E Todd.
\newblock Matching as an econometric evaluation estimator: Evidence from evaluating a job training programme.
\newblock \emph{The review of economic studies}, 64\penalty0 (4):\penalty0 605--654, 1997.

\bibitem[Hirano et~al.(2003)Hirano, Imbens, and Ridder]{hirano2003efficient}
Keisuke Hirano, Guido~W Imbens, and Geert Ridder.
\newblock Efficient estimation of average treatment effects using the estimated propensity score.
\newblock \emph{Econometrica}, 71\penalty0 (4):\penalty0 1161--1189, 2003.

\bibitem[Holland(1986)]{holland1986statistics}
Paul~W Holland.
\newblock Statistics and causal inference.
\newblock \emph{Journal of the American statistical Association}, 81\penalty0 (396):\penalty0 945--960, 1986.

\bibitem[Huber(2023)]{huber2023causal}
Martin Huber.
\newblock \emph{Causal analysis: Impact evaluation and Causal Machine Learning with applications in R}.
\newblock MIT Press, 2023.

\bibitem[Huntington-Klein(2021)]{huntington2021effect}
Nick Huntington-Klein.
\newblock \emph{The effect: An introduction to research design and causality}.
\newblock CRC Press, 2021.

\bibitem[Imai and Ratkovic(2014)]{imai2014covariate}
Kosuke Imai and Marc Ratkovic.
\newblock Covariate balancing propensity score.
\newblock \emph{Journal of the Royal Statistical Society: Series B (Statistical Methodology)}, 76\penalty0 (1):\penalty0 243--263, 2014.

\bibitem[Imbens(2004)]{imbens2004}
Guido~W Imbens.
\newblock Nonparametric estimation of average treatment effects under exogeneity: A review.
\newblock \emph{Review of Economics and Statistics}, pages 1--29, 2004.

\bibitem[Imbens(2015)]{imbens2015}
Guido~W Imbens.
\newblock Matching methods in practice: Three examples.
\newblock \emph{Journal of Human Resources}, 50\penalty0 (2):\penalty0 373--419, 2015.

\bibitem[Imbens and Rubin(2015)]{imbens2015causal}
Guido~W Imbens and Donald~B Rubin.
\newblock \emph{Causal Inference in Statistics, Social, and Biomedical Sciences}.
\newblock Cambridge University Press, 2015.

\bibitem[Imbens and Wooldridge(2009)]{imbens2009recent}
Guido~W Imbens and Jeffrey~M Wooldridge.
\newblock Recent developments in the econometrics of program evaluation.
\newblock \emph{Journal of economic literature}, 47\penalty0 (1):\penalty0 5--86, 2009.

\bibitem[Imbens et~al.(2001)Imbens, Rubin, and Sacerdote]{imbensrubinsacerdote}
Guido~W Imbens, Donald~B Rubin, and Bruce~I Sacerdote.
\newblock Estimating the effect of unearned income on labor earnings, savings, and consumption: Evidence from a survey of lottery players.
\newblock \emph{American Economic Review}, pages 778--794, 2001.

\bibitem[Kline(2011)]{kline2011oaxaca}
Patrick Kline.
\newblock Oaxaca-blinder as a reweighting estimator.
\newblock \emph{American Economic Review}, 101\penalty0 (3):\penalty0 532--537, 2011.

\bibitem[LaLonde(1986)]{LaLonde}
Robert~J LaLonde.
\newblock Evaluating the econometric evaluations of training programs with experimental data.
\newblock \emph{The American Economic Review}, pages 604--620, 1986.

\bibitem[Lechner(1999)]{lechner1999earnings}
Michael Lechner.
\newblock Earnings and employment effects of continuous gff-the-job training in east germany after unification.
\newblock \emph{Journal of Business \& Economic Statistics}, 17\penalty0 (1):\penalty0 74--90, 1999.

\bibitem[Lechner(2002)]{lechner2002program}
Michael Lechner.
\newblock Program heterogeneity and propensity score matching: An application to the evaluation of active labor market policies.
\newblock \emph{Review of Economics and Statistics}, 84\penalty0 (2):\penalty0 205--220, 2002.

\bibitem[Li et~al.(2018)Li, Morgan, and Zaslavsky]{li2018balancing}
Fan Li, Kari~Lock Morgan, and Alan~M Zaslavsky.
\newblock Balancing covariates via propensity score weighting.
\newblock \emph{Journal of the American Statistical Association}, 113\penalty0 (521):\penalty0 390--400, 2018.

\bibitem[Neyman(1923/1990)]{neyman1923}
Jerzey Neyman.
\newblock On the application of probability theory to agricultural experiments. essay on principles. section 9.
\newblock \emph{Statistical Science}, 5\penalty0 (4):\penalty0 465--472, 1923/1990.

\bibitem[Robins and Rotnitzky(1995)]{robins1995semiparametric}
James~M Robins and Andrea Rotnitzky.
\newblock Semiparametric efficiency in multivariate regression models with missing data.
\newblock \emph{Journal of the American Statistical Association}, 90\penalty0 (429):\penalty0 122--129, 1995.

\bibitem[Robins and Rotnitzky(2001)]{robins2001comment}
James~M Robins and Andrea Rotnitzky.
\newblock Comment on “inference for semiparametric models: Some questions and an answer” by pj bickel and j. kwon.
\newblock \emph{Statistica Sinica}, 11:\penalty0 920--936, 2001.

\bibitem[Robins et~al.(1994)Robins, Rotnitzky, and Zhao]{robins1994estimation}
James~M Robins, Andrea Rotnitzky, and Lue~Ping Zhao.
\newblock Estimation of regression coefficients when some regressors are not always observed.
\newblock \emph{Journal of the American statistical Association}, 89\penalty0 (427):\penalty0 846--866, 1994.

\bibitem[Rosenbaum(1984)]{rosenbaum1984consequences}
Paul~R Rosenbaum.
\newblock The consequences of adjustment for a concomitant variable that has been affected by the treatment.
\newblock \emph{Journal of the Royal Statistical Society Series A: Statistics in Society}, 147\penalty0 (5):\penalty0 656--666, 1984.

\bibitem[Rosenbaum and Rubin(1983{\natexlab{a}})]{rosenbaum1983central}
Paul~R Rosenbaum and Donald~B Rubin.
\newblock The central role of the propensity score in observational studies for causal effects.
\newblock \emph{Biometrika}, 70\penalty0 (1):\penalty0 41--55, 1983{\natexlab{a}}.

\bibitem[Rosenbaum and Rubin(1983{\natexlab{b}})]{rosenbaumrubin1983assessing}
Paul~R Rosenbaum and Donald~B Rubin.
\newblock Assessing sensitivity to an unobserved binary covariate in an observational study with binary outcome.
\newblock \emph{Journal of the Royal Statistical Society. Series B (Methodological)}, pages 212--218, 1983{\natexlab{b}}.

\bibitem[Rosenbaum et~al.(1987)]{rosenbaum1987role}
Paul~R Rosenbaum et~al.
\newblock The role of a second control group in an observational study.
\newblock \emph{Statistical Science}, 2\penalty0 (3):\penalty0 292--306, 1987.

\bibitem[Rubin(1973)]{rubin1973use}
Donald~B Rubin.
\newblock The use of matched sampling and regression adjustment to remove bias in observational studies.
\newblock \emph{Biometrics}, pages 185--203, 1973.

\bibitem[Rubin(1974)]{rubin1974estimating}
Donald~B Rubin.
\newblock Estimating causal effects of treatments in randomized and nonrandomized studies.
\newblock \emph{Journal of educational Psychology}, 66\penalty0 (5):\penalty0 688, 1974.

\bibitem[Rubin(1978)]{rubin1978bayesian}
Donald~B Rubin.
\newblock Bayesian inference for causal effects: The role of randomization.
\newblock \emph{The Annals of statistics}, pages 34--58, 1978.

\bibitem[Rubin(2006)]{rubin2006matched}
Donald~B Rubin.
\newblock \emph{Matched sampling for causal effects}.
\newblock Cambridge University Press, 2006.

\bibitem[Smith and Todd(2001)]{smith2001reconciling}
Jeffrey~A Smith and Petra~E Todd.
\newblock Reconciling conflicting evidence on the performance of propensity-score matching methods.
\newblock \emph{American Economic Review}, 91\penalty0 (2):\penalty0 112--118, 2001.

\bibitem[Smith and Todd(2005)]{smith2005does}
Jeffrey~A Smith and Petra~E Todd.
\newblock Does matching overcome lalonde's critique of nonexperimental estimators?
\newblock \emph{Journal of econometrics}, 125\penalty0 (1-2):\penalty0 305--353, 2005.

\bibitem[Van~der Laan and Rose(2011)]{van2011targeted}
Mark~J Van~der Laan and Sherri Rose.
\newblock \emph{Targeted learning: causal inference for observational and experimental data}.
\newblock Springer Science \& Business Media, 2011.

\bibitem[Wager(2024)]{wager2024}
Stefan Wager.
\newblock \emph{Causal Inference: A Statistical Learning Approach}.
\newblock Cambridge University Press, 2024.

\bibitem[Wager and Athey(2015)]{wagerathey}
Stefan Wager and Susan Athey.
\newblock Causal random forests.
\newblock \emph{arXiv preprint}, 2015.

\bibitem[Wager and Athey(2017)]{wager2017estimation}
Stefan Wager and Susan Athey.
\newblock Estimation and inference of heterogeneous treatment effects using random forests.
\newblock \emph{Journal of the American Statistical Association}, \penalty0 (just-accepted), 2017.

\bibitem[Wager and Athey(2018)]{wager2018estimation}
Stefan Wager and Susan Athey.
\newblock Estimation and inference of heterogeneous treatment effects using random forests.
\newblock \emph{Journal of the American Statistical Association}, 113\penalty0 (523):\penalty0 1228--1242, 2018.

\bibitem[Xu(2023)]{xu2023causal}
Yiqing Xu.
\newblock Causal inference with time-series cross-sectional data: A reflection.
\newblock In \emph{Oxford Handbook of Engaged Methodological Pluralism}, chapter~30. Oxford University Press, November 2023.

\bibitem[Zhao and Percival(2017)]{zhao2017entropy}
Qingyuan Zhao and Daniel Percival.
\newblock Entropy balancing is doubly robust.
\newblock \emph{Journal of causal inference}, 5\penalty0 (1):\penalty0 20160010, 2017.

\bibitem[Zubizarreta(2015)]{zubizarreta2015stable}
Jose~R. Zubizarreta.
\newblock Stable weights that balance covariates for estimation with incomplete outcome data.
\newblock \emph{Journal of the American Statistical Association}, 110\penalty0 (511):\penalty0 910--922, 2015.
\newblock \doi{10.1080/01621459.2015.1023805}.

\bibitem[Zubizarreta et~al.(2023)Zubizarreta, Stuart, Small, and Rosenbaum]{zubizarreta2023handbook}
Jos{\'e}~R Zubizarreta, Elizabeth~A Stuart, Dylan~S Small, and Paul~R Rosenbaum.
\newblock \emph{Handbook of Matching and Weighting Adjustments for Causal Inference}.
\newblock CRC Press, 2023.

\end{thebibliography}

\clearpage
\appendix
\onehalfspacing
\setcounter{page}{1}
\setcounter{table}{0}
\setcounter{figure}{0}
\setcounter{equation}{0}
\setcounter{footnote}{0}
\renewcommand\thetable{A\arabic{table}}
\renewcommand\thefigure{A\arabic{figure}}
\renewcommand{\thepage}{A-\arabic{page}}
\renewcommand{\theequation}{A\arabic{equation}}
\renewcommand{\thefootnote}{A\arabic{footnote}}

\vspace{0em}
\section{Tables 4-6 in \citet{LaLonde}}
\bigskip

The following tables are adapted from Tables 4, 5, and 6 in \citet{LaLonde}. We thank Robert LaLonde's estate for allowing us to include these tables in the Appendix.

\begin{table}[!ht]
\includegraphics[scale = 0.4]{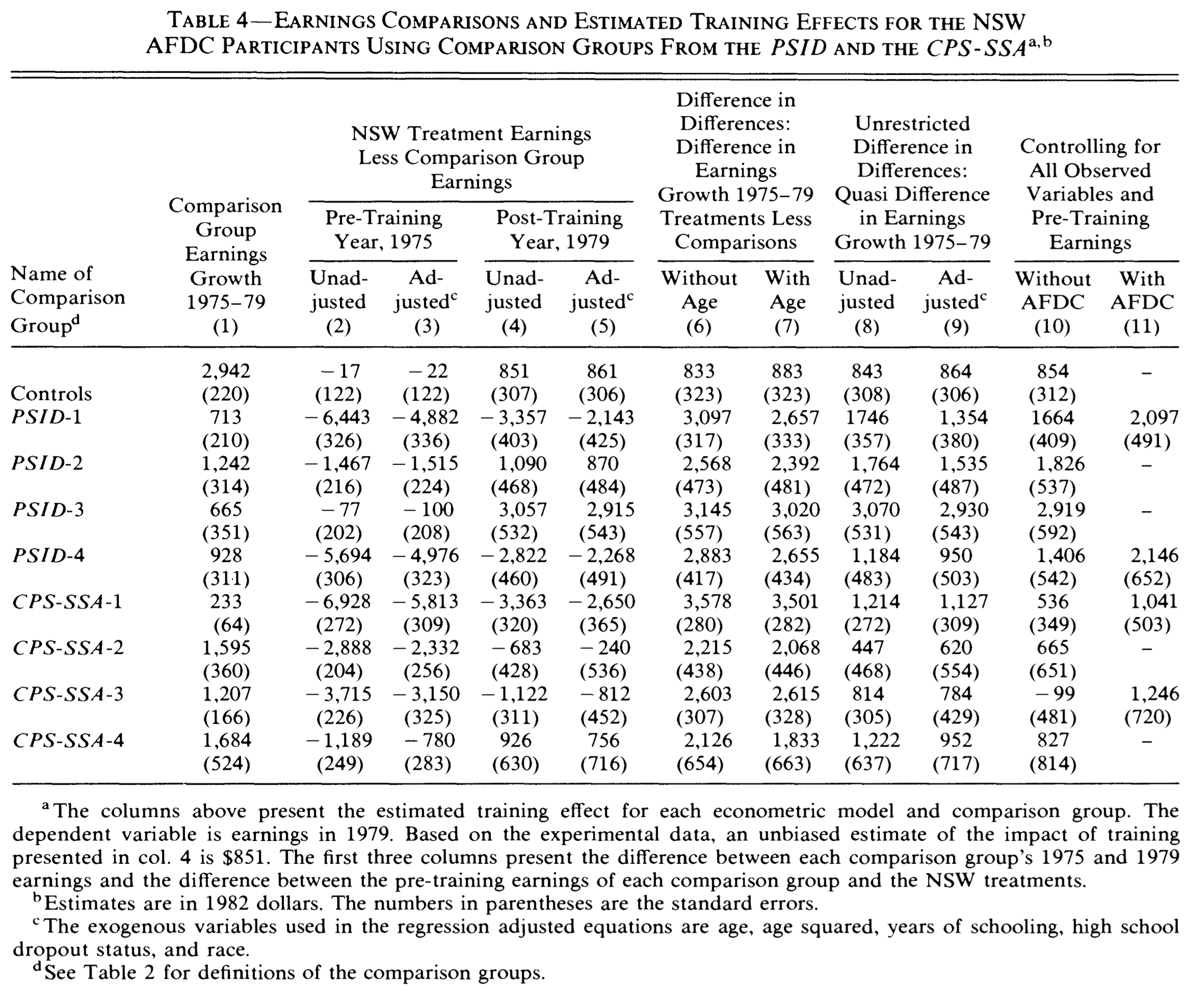}
\end{table}
\clearpage

\begin{table}[!ht]
\includegraphics[scale = 0.4]{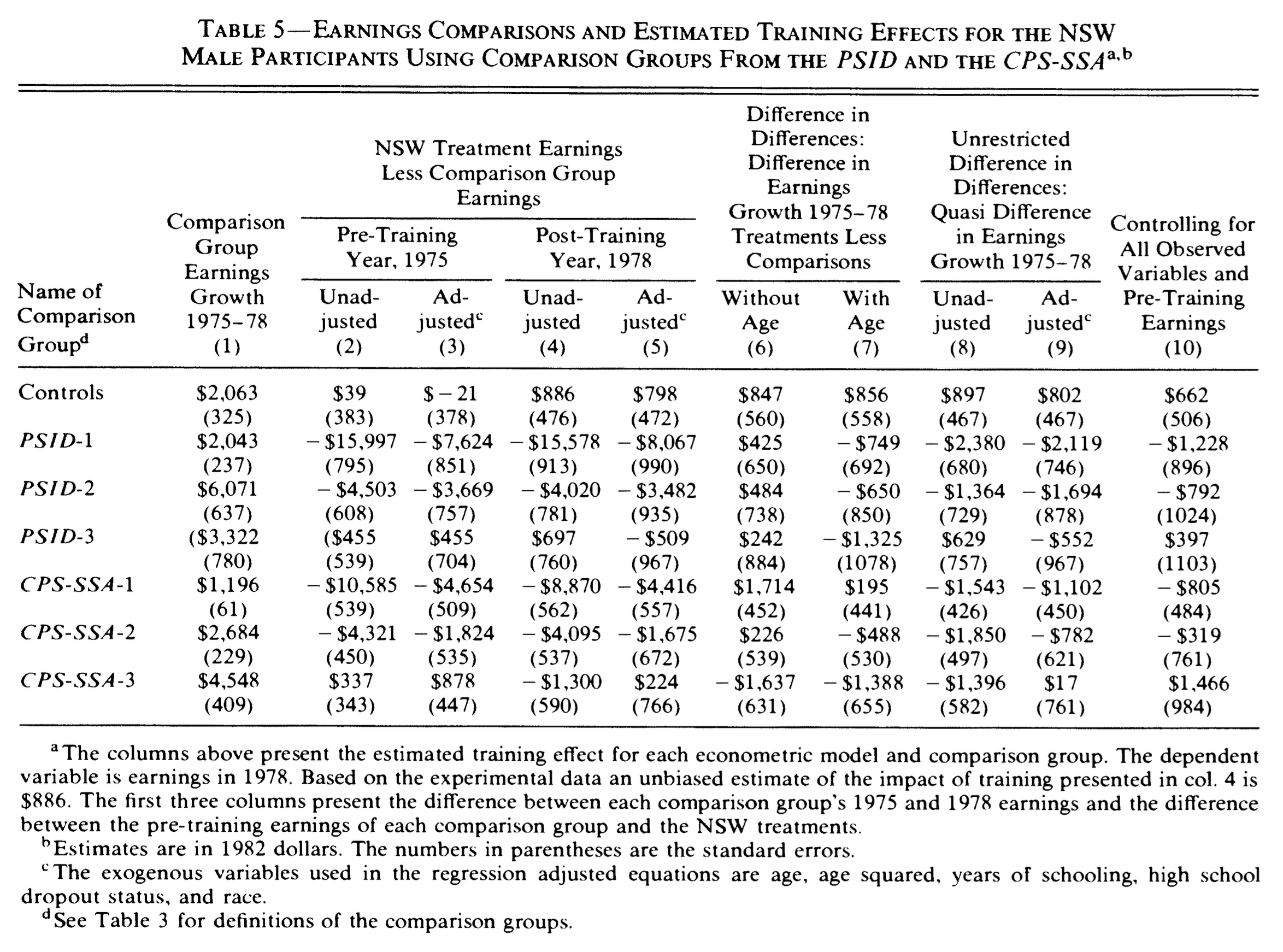}
\end{table}
\clearpage

\begin{table}[!ht]
\includegraphics[scale = 0.4]{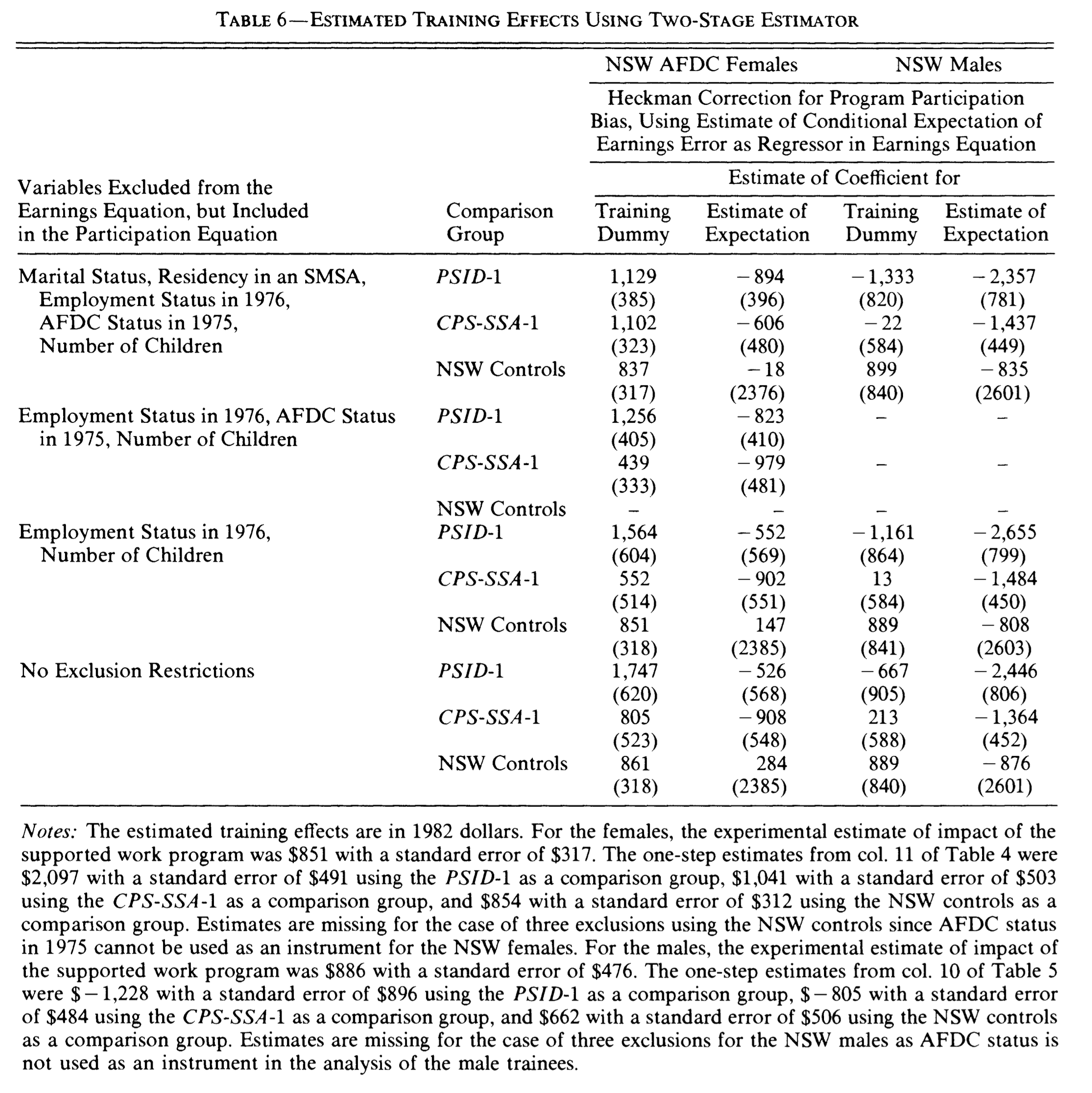}
\end{table}

\end{document}